\documentclass[preprint2]{aastex}

\tighten 

\input{psfig.sty}

\shorttitle{H$\alpha$ Atlas of Virgo Cluster Galaxies}
\shortauthors{Koopmann et al.}

\begin{document}

\title{An Atlas of H$\alpha$ and R Images and Radial Profiles
of 63 Bright Virgo Cluster Spiral Galaxies}

\author{Rebecca A. Koopmann}
\affil{Union College}
\affil{Department of Physics, Schenectady, NY 12308}
\email{koopmanr@union.edu}

\author{Jeffrey D. P. Kenney}
\affil{Astronomy Department}
\affil{Yale University, P.O. Box 208101, New Haven, CT 06520-8101}
\email{kenney@astro.yale.edu}

\author{Judith Young}
\affil{University of Massachusetts}
\affil{Five Colleges Radio Observatory}
\affil{6191 Graduate Res. Ctr., Amherst, MA, 01003}
\email{young@phast.umass.edu}

\begin{abstract}
Narrow-band H$\alpha$ and broadband R images and radial profiles are presented
for 63 bright spiral galaxies in the Virgo Cluster. 
The sample is complete for Sb-Scd
galaxies with B$_T^0$ $\le$ 12 and inclination $\le$ 75$^{\circ}$.
Isophotal radii, disk scalelengths, 
concentration parameters, and integrated fluxes are derived for the sample
galaxies. 

\end{abstract}


\section{Introduction}
How has the environment affected the star formation properties 
of spiral galaxies in clusters? Previous measurements 
of the global star formation rates at H$\alpha$, UV, and far-infrared 
wavelengths of cluster spirals show that some cluster spirals have
reduced global star formation (Kennicutt 1983, Bicay \& Giovanelli 1987, 
Kodaira et al.
1990, Moss \& Whittle 1993), others have similar global star formation
rates (Kennicutt, Bothun, \& Schommer 1984,
Gavazzi et al. 1991, Donas et al. 1990), and some have enhanced star formation
rates (Moss \& Whittle 1993, Bennett \& Moss 1998)
with respect to field counterparts. 

Studies of the spatial distributions of star formation have been more
limited and have usually concentrated on field spiral samples (e.g., Hodge
\& Kennicutt 1983; Ryder 
\& Dopita 1993; Phillips 1993;
Gonz\'{a}lez Delgado et al. 1997; Hameed \& Devereux 1999).
Based on these studies, star formation tends to
be distributed over much of the optical disk, often concentrated along 
spiral arms and in rings. 
To date there has been little comparison of high resolution
spatial distributions of star formation for large samples of cluster
and field galaxies. Such comparisons could help reveal where in the disk
star formation has been reduced or enhanced in cluster galaxies, leading
to strong constraints on the environmental mechanisms which transform
cluster galaxies.

The R and H$\alpha$ images presented in this paper are part of an 
extensive study of the spatial distributions of massive star formation in 
nearly
100 Virgo Cluster and isolated spiral and lenticular galaxies. 
The main goal of the work is to
better understand the relative distributions of star formation in cluster
and isolated spirals and to probe the role of the environment in the 
star formation properties of cluster spirals. 

The sample presented in this paper is drawn from the nearby Virgo Cluster
of galaxies. The Virgo
Cluster is an excellent laboratory for studies of star formation histories
since it is near enough that galaxy morphology can be examined in detail.
It has also been surveyed at many different wavelengths, including aperture 
H$\alpha$ (Kennicutt 1983), HI (Warmels 
1988, Cayatte et al. 1990, Hoffman et al. 1989), CO (Kenney \& Young 1989),
near-IR (Boselli et al. 1997), radio continuum (Kotanyi 1980) and X-ray 
(Boehringer et al. 1994). 
The structure and membership of the cluster are well-studied 
(Binggeli, Sandage, \& Tammann 1985; Binggeli, Tammann, \& Sandage 1987; 
Binggeli, Popescu, \& Tammann 1993; 
Yasuda, Fukugita, \& Okamura  1997; Gavazzi et al.
1999, Schindler, Binggeli, \& Boehringer 1999). Because the cluster has 
not yet relaxed, there is a enhanced likelihood of detecting
ongoing environmental alteration. 

This paper presents continuum-subtracted H$\alpha$+[N II] and broadband
R images of 63 bright spiral galaxies in the Virgo Cluster. 
R and H$\alpha$ profiles are derived from surface photometry of the images. 
Isophotal radii, disk scalelengths, integrated fluxes, and concentration
parameters are measured from the radial profiles. 
Total H$\alpha$ fluxes derived from 50 of the images presented here
have already been published in Young et al. (1996), although the
calibration has been adjusted in this paper.

This study marks the first presentation of H$\alpha$ spatial distributions 
for a large and highly complete sample of Virgo Cluster spirals. (Previous
results based on H$\alpha$
images and surface brightnesses for several Virgo Cluster spirals are
found in Kennicutt 1989) In addition, this study presents R surface 
photometry, which
has been relatively rare for Virgo Cluster spirals. (Some R surface 
photometry is available for several galaxies
within large surveys, e.g., de Jong \& van der Kruit 1994). In fact, even 
total
R magnitudes have previously been published for only about half of this 
sample (Pierce \& Tully 1988; Schroeder \& Visvanathan 1996). 

This paper is part of a series.
Images and radial profiles of 30 isolated spiral galaxies are presented
in Koopmann \& Kenney (2001a). 
Concentration parameters derived from the
R radial profiles and H$\alpha$ fluxes normalized by the R fluxes are used
in Koopmann \& Kenney (1998) to show that the Hubble 
classification scheme
is inadequate in describing the morphology of at least 25\% of Virgo Cluster
spirals. The radial distributions of R and H$\alpha$ light of the Virgo
Cluster and isolated galaxies are analyzed and compared
in Koopmann \& Kenney (2001b).

The selection of the Virgo cluster galaxies is described in 
Section~\ref{sample}.
Sections~\ref{obs} and ~\ref{reduc} describe the observations and 
reduction procedures. The images and derived radial profiles
are presented in Section~\ref{images}.
Surface photometry procedures are outlined in 
Section~\ref{surfphot}.

\section{Sample Selection}
\label{sample}
\subsection{Structure of the Virgo Cluster}
\label{vstruct}
Studies of the substructure of the Virgo
Cluster (including Tully \& Shaya 1984; Pierce \& Tully 1988; Binggeli
et al. 1987, 1993; Yasuda et al. 1997; and Gavazzi et al. 1999, Schindler
et al. 1999) agree that
the Virgo Cluster is made up of several associated clouds or groups of 
galaxies, with 
two main components: cluster A, associated with M87, and
cluster B, associated with M49 (although Gavazzi et al. 1999
place M49 closer than cluster B, based on Tully-Fisher distances). 
Within subcluster A is a smaller core of mostly spheroidal galaxies. 
Even this core is not in dynamical equilibrium (Binggeli et al. 1987, 1993),
indicating the dynamical youth of the cluster. 
The rest of the cluster has a filamentary structure, 
some of which is extended at least partially along the line of sight. 
One filament is the Southern Extension region of the cluster, which is a  
cloud of galaxies located adjacent to and at about the same distance
as the central components (Binggeli et al. 1993).
Three clouds of galaxies, the W (de Vaucouleurs 1961), W', and M 
(Ftaclas et al. 1984) clouds, are projected on the cluster; 
studies consistently place these clouds
in the near-background at about two times the distance of the
cluster.

\begin{figure}
\centerline{\includegraphics[width=5.5in]{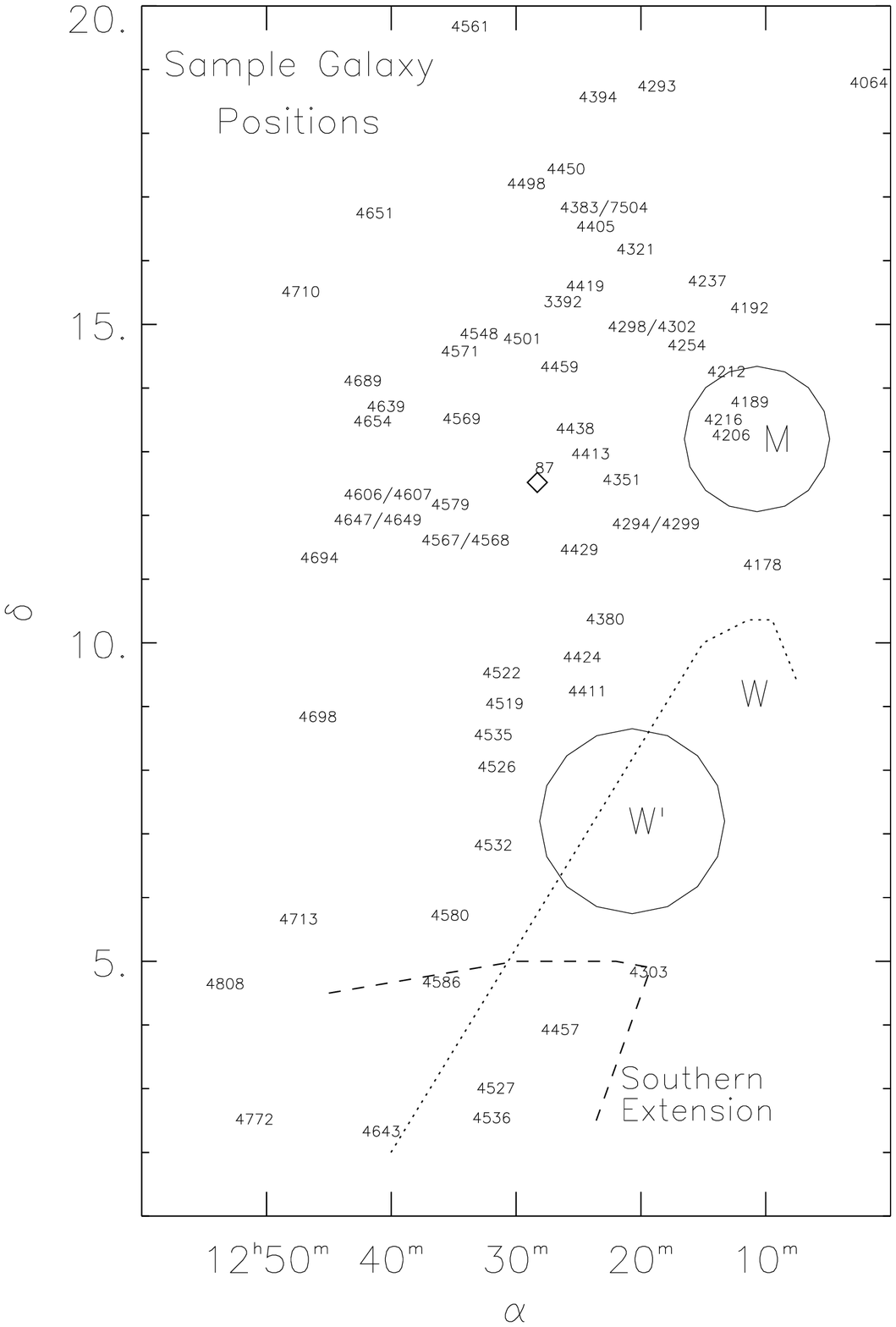}}
\caption{Locations of sample galaxies in the Virgo Cluster. Galaxy positions
are symbolized by the NGC/IC number of the galaxy. The projected
locations of the M, and W' clouds (circles), the W cloud (dotted line), and the
Southern Extension (dashed
lines) are indicated, based on Figure 4 from Binggeli et al. (1993). 
The position of M87 is indicated by the diamond. \label{virloc}}
\end{figure}

Membership information from the Virgo Cluster Catalog 
(Binggeli et al. 1985; hereafter BST) 
was considered in the selection of sample galaxies. Figure~\ref{virloc} 
shows a map of the locations of sample galaxies, with the projected
regions of different clouds indicated. Sample galaxies 
are located throughout the cluster, with a range of 1-10$^{\circ}$ 
projected distance from M87. Possible members of the
W, W', and M clouds were avoided; however the data of Binggeli et al. (1993) 
suggest that 4 galaxies in our sample may be located in these clouds
(NGC 4189, NGC 4303, NGC 4527, and NGC 4536). Several galaxies may be
members of the Southern Extension (including NGC 4457 and NGC 4586). 
Eight galaxies which
lie outside of the limited survey region of BST were also observed (NGC 4064,
NGC 4561, NGC 4643,  NGC 4651, NGC 4710, NGC 4713, NGC 4772, NGC 4808);
Tully-Fisher distance estimates for many of these galaxies indicate a
distance consistent with the Virgo Cluster (Kenney 1987, Rubin et al. 1999).

We assume a distance of 16 
Mpc for the Virgo Cluster (Jacoby et al. 1992; Yasuda
et al. 1997; Kelson et al. 2000). 
Most of our analyses were designed to be 
distance-independent, since we will compare to a sample of isolated galaxies
of different distances. Hence the possibility of a significant depth of
the Virgo Cluster will not influence most of our conclusions.

\subsection{Properties of Sample Galaxies}
The Virgo Cluster sample galaxies range in type from S0-Sm.
The galaxies were chosen to have B$_T^0$ brighter than 13.5 (corresponding
to M$_B$ of -17.5 at an assumed distance of 16 Mpc) and 
inclinations mostly 
$<$ 75$^{\circ}$. Seven of the 63 galaxies have higher inclination; some
of these were observed in the same field of view as target galaxies.
Observations of 50 of the galaxies
were obtained as part of a study of star formation rates and efficiencies in
120 spiral galaxies (Young et al. 1996). This study
had as selection criteria
IRAS flux cutoffs of S$_{60}$ $>$ 5 Jy or S$_{100}$ $>$ 10 Jy. 
Thirteen additional galaxies,
mostly of cataloged Hubble type S0-Sa, 
were observed as part of the current study and were not constrained by the
IRAS flux criteria.

The properties of the galaxies observed are listed in Table 1. 
Due to the uncertain nature of some
Virgo cluster morphological classifications, Table 1 provides Hubble types from
two major classifiers. Classifications were taken from 
the Revised Shapley-Ames Catalog of Bright Galaxies (Sandage \&
Tamann 1987, hereafter RSA), BST, 
and the Third Reference Catalog of Bright
Galaxies (deVaucouleurs et al. 1991; hereafter RC3).
Note that several galaxies have mixed classifications
like Sc/Sa and that there are disagreements between the two major
classifiers for a number of galaxies. The important
issue of morphological classification
is addressed further in Koopmann \& Kenney (1998, 2001b).

\placetable{tab1}

55 out of 63 galaxies are located within the region of the Virgo Cluster 
surveyed by BST. 
The completeness values for 46 galaxies within the BST limits with
B$_T^0$ $<$ 13 and inclination $<$ 75$^{\circ}$ are presented in 
Table~\ref{tab2} 
for different Hubble types. For simplicity, Hubble types were taken at
face value from BST and include all peculiar types. 
Three galaxies with Sc/Sa or
Sc/S0 classifications were counted as Sc galaxies. (We describe the
morphologies of these galaxies and analyze their properties separately
in other papers in this series.) Seven of the sample galaxies within the 
BST limits have 
inclinations $\ge$ 75$^{\circ}$ and 2 Sc galaxies have B$_T^0$ between 13 and 
13.5. Except for
one Sa galaxy, the Virgo sample within the BST survey region
is complete to B$_T^0$ $<$ 12 for types Sa-Sm.

\placetable{tab2}

\section{Observations}
\label{obs}
Observations of sample galaxies were obtained 
at the KPNO (\#1) 0.9-m,  the CTIO 0.9-m, and the 3.5-m
WIYN telescopes between March 1987 and February 1997. All observations
were calibrated under photometric conditions. 
Exposure times ranged from 3-15 min in 
an R filter (nearly Mould or standard Harris/Kron-Cousins R), and 1-2 hrs in an
H$\alpha$ filter of width 60-80 \AA, centered either near 6563 \AA \
or near 6600 \AA \  (for galaxies with v $>$ 1000 km/s). 

\begin{figure}
\centerline{
\includegraphics[height=4in,angle=270]{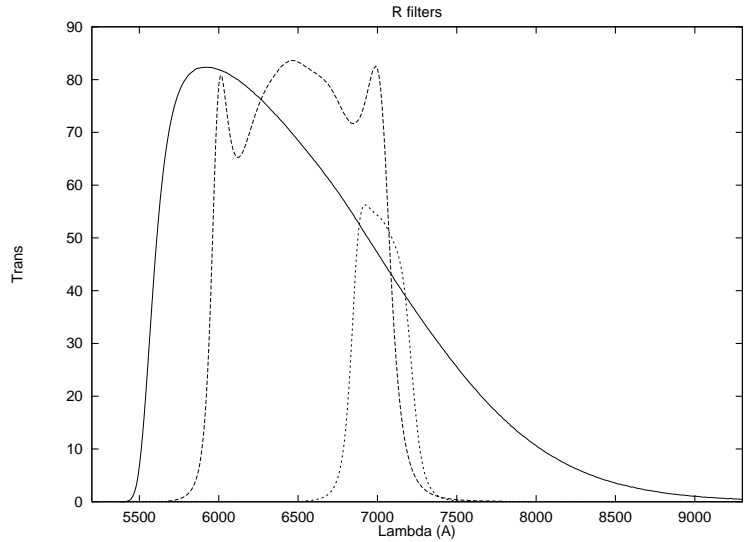}
}
\caption{Red passband filter curves: Harris R 
(solid), nearly Mould R (dashed), and narrow R (dotted). Filters are
respectively abbreviated as R, nmR, and sR in Table~\ref{obslog}. \label{filt}}
\end{figure}

The observations of the Virgo sample galaxies were made through
several different red filters. 
Between 1992 and 1997, images were obtained using a
standard Harris (Kron-Cousins) R filter ($\lambda$ = 6425 \AA, 
$\Delta\lambda$ = 1500 \AA). Images obtained earlier were
observed in either a nearly-Mould R filter ($\lambda$ = 6470 \AA, 
$\Delta\lambda$ = 1110 \AA) or, for 12\% of the sample, a narrow red filter
($\lambda$ = 7024 \AA, $\Delta\lambda$ = 380 \AA).  
Filter curves are shown in Figure~\ref{filt}.

In order to compare these observations to an isolated comparison sample,
observations in nearly Mould R were transformed to Harris R (see 
Section~\ref{fc}), using observations of Landolt standard stars in 
both filters. Galaxies observed in the narrow red filter were
reobserved in Harris R in February 1997 at WIYN and the CTIO 0.9-m telescopes.

An observing log for the galaxies is presented in Table~\ref{obslog}. 
Galaxies are
listed in order of NGC/IC number, with two to three lines describing the
R and H$\alpha$ observations. 

\placetable{obslog}
\placetable{chip}
\placetable{filter}

A number of CCD chips were used over the course of the observations.
Details of the chips are listed in Tables~\ref{obslog} and~\ref{chip}.
Characteristics of the filters used in the observations are given in
Table 5.
The fields-of-view of the observations were large enough to image each galaxy
to at least an isophotal radius of 24 mag arcsec$^{-2}$
and, in most cases, sufficient blank sky to accurately estimate the
sky background of the chip. R observations of several of the larger
Virgo galaxies obtained in the late 1980's were repeated with larger
format CCD's in the 1990's to obtain a larger field-of-view.

Observations of
the spectrophotometric standard stars from the lists of Massey et al. (1988)
and Hamuy et al. (1992) were made
for the calibration to absolute flux. Most observations obtained between 
1992 and 1997 included
exposures of Landolt standards (1992a) to derive the extinction coefficient.

\section{Reduction Procedures}
\label{reduc}
\subsection{Processing}
All images were bias-subtracted and flat-fielded using 5-10 dome-flats and/or
twilight sky flats. For images taken at the CTIO 0.9-m, illumination 
corrections derived from twilight sky flats applied to the dome-flattened 
images significantly improved the flatness
of the R sky background (from $\sim$ 2-3\% to $<$1\%). 
Standard IRAF (Tody 1993) tasks were used for the processing steps.

For some galaxies, 2-3 exposures were obtained. The images
were registered and aligned using the centroids of bright stars, and
averaged using the \it combine \rm task in IRAF, with cosmic ray rejection
based on the noise parameters of the CCD.
In the case of seeing differences, images were convolved with a Gaussian
function to the worst
seeing before combining. For these images and single exposure 
images, the \it bclean \rm task in the software package FIGARO and the
\it imedit \rm task in IRAF were used to remove cosmic rays. 

\subsection{Absolute Flux Calibration}

Extinction coefficients were derived from 
exposures of Landolt standards where
possible. Otherwise the standard extinction curve at KPNO or CTIO was used.
Most of the R and all the H$\alpha$ 
images were flux calibrated based on spectrophotometric standard stars
(Massey et al. 1988, Hamuy et al. 1992). 

For most of the observations, the filter response curve was convolved with the 
spectrum of the standard to compute the expected flux within the filter. 
This approach is
especially important for flux calibration of images in the Harris R
filter. The H$\alpha$ calibrations prior to 1992 were made by approximating the
total flux expected in the filter by integrating the magnitude 
of the standard at the central wavelength of the filter over the filter band
width. 
This approximation has an accuracy on the order of 1-2\% for the narrow-band
H$\alpha$ filter, well below
the error contributed from other sources. This approximation is not
appropriate for the broadband R filters, particularly the Harris R filter,
due to its complex transmission curve. 

\subsubsection{Reducing to a Standard System: R}
\label{fc}
Because we wish to compare R profiles, fluxes, and size scales within the
Virgo sample and to the comparison isolated sample (Koopmann \& Kenney 2001a), 
standardization to the Harris R system is necessary. 
Galaxies observed in nearly Mould R were transformed to Harris R. Because no 
transformation equation was available in the
literature, one was derived through observations of a series of Landolt (1992a)
standard stars in both filters. The transformation was well described by
a simple multiplicative factor, with no color dependence:
$$\rm \frac{F_R}{F_{mR}} = 1.35 \pm 0.05$$
where F$_R$ is the flux in R and F$_{nmR}$ 
is the flux in the nearly Mould R filter.
However, the scatter in the multiplicative factor introduces an additional
4\% error in the R fluxes of galaxies observed in the nearly Mould R filter.

All quantities derived from the R photometry presented in this and succeeding
papers are calibrated to the Harris R system.
R flux values can be converted to R magnitudes using a zero point
of 13.945, which was derived from comparison of spectrophotometric standard 
star absolute fluxes and magnitudes listed by Landolt (1992b). 
This relationship was also used to flux calibrate two R images for which 
spectrophotometric standards were not available. 

\begin{figure}
\centerline{\includegraphics[height=5in,angle=90]{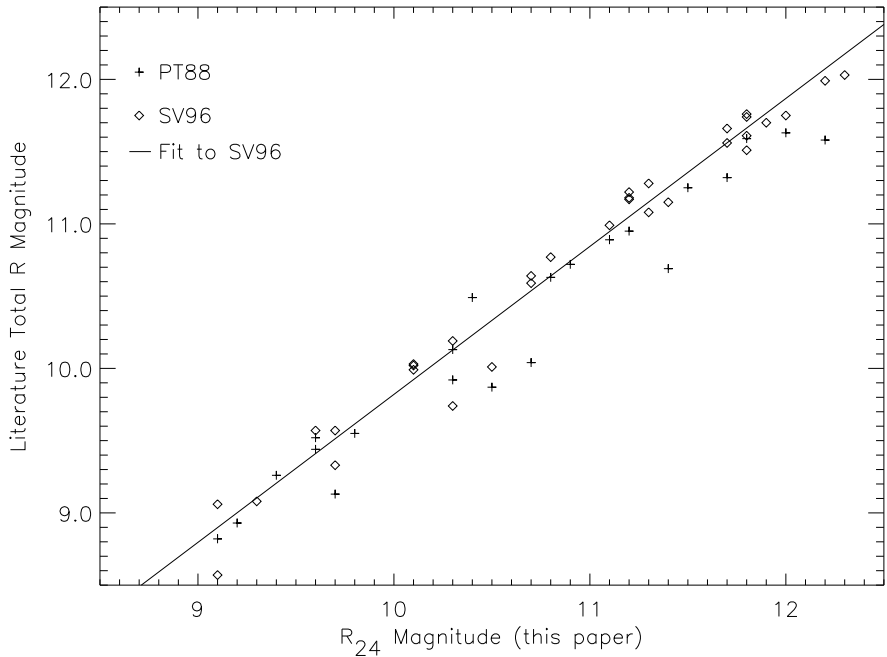}}
\caption{Comparison between total R magnitudes from Pierce \& Tully (1988)
and Schroeder \& Visvanathan (1996) and the R$_{24}$ isophotal magnitudes for 
galaxies
in common. The solid line shows a least-squares fit to the 
Schroeder \& Visvanathan
data, with a relation R$_{tot}$ = 1.02 R$_{24}$ - 0.04.  
There is a 3\% uncertainty in the slope.
The Pierce \& Tully data has a larger scatter, with
a 5\% uncertainty in the slope of a linear fit.
\label{rphotcomp}}
\end{figure}

There are few previous measurements of R surface photometry for Virgo
Cluster galaxies. Total R magnitudes are presented by
Pierce and Tully (1988) for 22 galaxies in our sample, and by Schroeder \&
Visvanathan(1996)
for 33 galaxies in our sample. 
We find a correlation between the total R magnitudes
and our isophotal (to 24 R mag arcsec$^{-2}$) magnitudes for both
sets of data, as shown in Figure~\ref{rphotcomp}.
The Pierce \& Tully magnitudes are 0.25
magnitude higher in the median than our isophotal magnitudes, with
a 5\% uncertainty in the slope of a linear correlation. The Schroeder \&
Visvanathan
magnitudes are 0.11 magnitude higher in the median, with a 3\% uncertainty
in the slope of a linear correlation.

\subsubsection{Reducing to a Standard System: H$\alpha$}
The H$\alpha$ images are calibrated to absolute flux units.
No correction was made for contamination by the 2 [N II] lines 
($\lambda$ $\lambda$=6548.1, 6583.8 \AA), which  
also lie within the filter bandpass. This assumption is based on the
work of Kennicutt \& Kent (1983) and Kennicutt (1992), who evaluate
the relative [N II] contribution to the flux in H II regions. They find
a  median [N II]/H$\alpha$ of 0.53, with
a scatter that is
significantly smaller than the scatter in overall H$\alpha$ emission strengths
in spiral galaxies.
Hereafter we will 
abbreviate H$\alpha$+[N II] as H$\alpha$.

H$\alpha$ observations prior to 1992 were originally calibrated using the
Barnes and Hayes (1984) Standard Star Manual. The fluxes based on
this system appear in Young et al. (1996). In this paper, we adjust the
calibration to the standard star measurements of 
Massey et al. (1988) for
consistency between different observations and our isolated sample
(11-20\% lower in flux depending on standard and filter).

\begin{figure}
\centerline{\includegraphics[height=6in,angle=90]{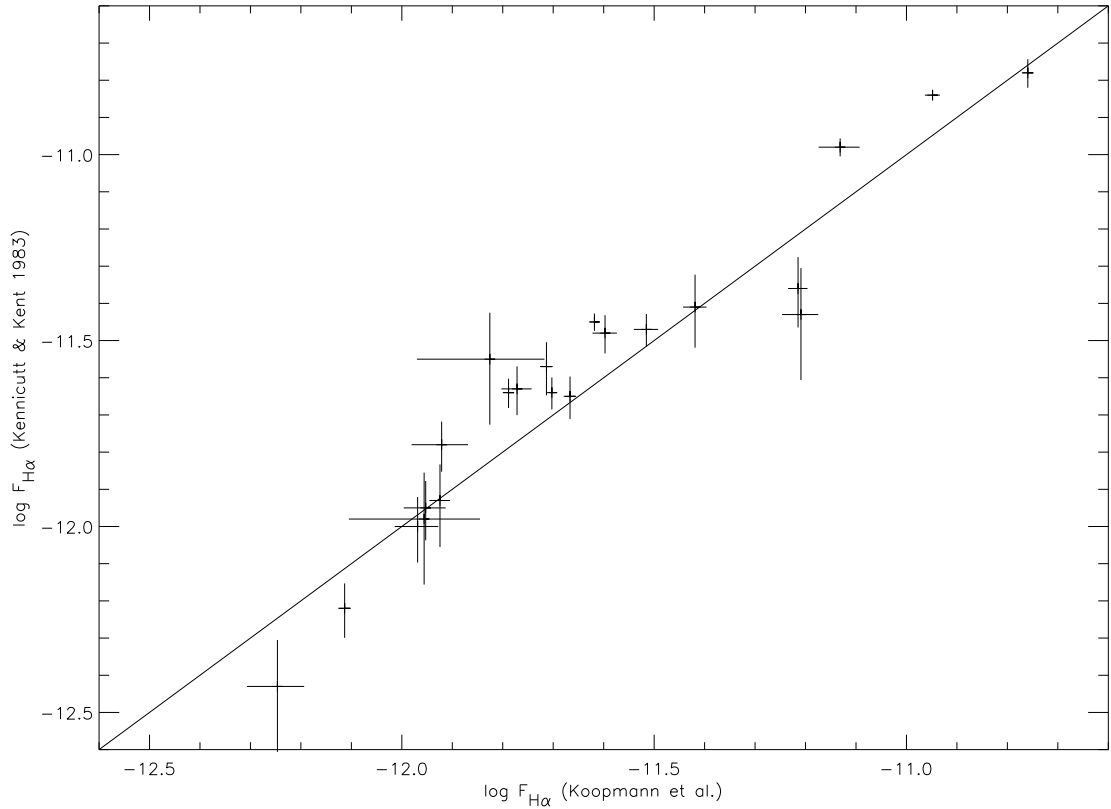}}
\caption{Comparison between H$\alpha$ fluxes measured by Kennicutt \& Kent
(1983) and this paper for 23 galaxies in common. The solid line shows a
1:1 correspondence. The error bars for our sample are calculated by adding
in quadrature the contributions from an assumed 2\% continuum subtraction
error and the error in the total flux due to the uncertainty in the sky
background value. 
\label{koopkenn}}
\end{figure}

The measured H$\alpha$ fluxes are well correlated with those measured by
Kennicutt \& Kent (1983) for 23 galaxies in common, 
as shown in Figure~\ref{koopkenn}. The solid line shows a one-to-one
correlation. There is a considerable scatter, with some individual galaxies
deviant by $>$ 30\%. However, most galaxies agree within 30\% and the
mean difference for the whole sample is close to 0. The differences for
individual galaxies are most likely due to uncertainties in the 
continuum subtraction 
background sky levels, as discussed in Sections~\ref{skyerr} 
and~\ref{hacssect}.

\subsection{Sky Background Subtraction}
\label{skyerr}
The most important source of error in the broadband radial profiles of the 
outer
disks of galaxies is the uncertainty in the sky background. 
It is therefore critical to 
obtain as flat a sky background as possible and carefully measure the
uncertainty in the sky level.
The standard deviation in the background and the uncertainty in the
overall sky level were determined by measuring
4-6 blank regions of the sky,
and by examining line and column plots, which were especially useful in
detecting large scale gradients. Standard deviations and uncertainties in
the overall sky level are listed in Table 3 in units of $10^{-18}$ erg 
cm$^{-2}$ s$^{-1}$ (a zeropoint of 13.945 may be used to convert the R
values to magnitudes).

Occasionally a planar gradient remained in the sky background
after processing, particularly in images obtained during bright time.
Where necessary, and where adequate blank sky was available,
a planar surface was fit (using \it imsurfit \rm in iraf) to the sky 
background 
and subtracted from the image to improve sky flatness. Where higher order 
gradients remained, usually due to improper flat fields or
scattered light, the effect was incorporated into the sky background error.

The uncertainty in the R sky is typically 0.5-2\%, 
except for cases in which the galaxy fills much of the
CCD frame. The sky background error 
tends to be higher in the narrow-band H$\alpha$, ranging from 1-6\%.

\subsection{H$\alpha$ Continuum Subtraction}
\label{hacssect}
Waller (1990) describes the use of broadband filters to continuum 
subtract narrow-line observations. We follow his procedure.

Before continuum subtraction, the R and H$\alpha$ images were background
subtracted, aligned, and convolved to the same FWHM. 
The R image was scaled to the H$\alpha$
image based on aperture photometry of 5-15 foreground stars. This scale
factor often needs adjustment, especially since the foreground field stars 
and the galaxy may be different in color. Adjustments were made
iteratively until a satisfactory subtraction was obtained over the
majority of the galaxy disk. Uncertainties in the scale factor and therefore
in the estimate of the continuum level 
are 2-5\%. Although these numbers appear small, they can result in a much
higher uncertainty in the measured H$\alpha$ flux or surface
brightness, since the continuum overwhelms the H$\alpha$ emission (see
Section~\ref{errors}.) An
additional source of uncertainty is the assumption of a constant scale
factor across the galaxy, especially if there are significant color changes 
caused by changes in stellar populations across the disk. Assumption of a
constant scale factor for the central regions and the disk is
particularly problematic (see Rand 1996). 
Often central regions have negative residuals
when the disk is well fit. Examples in this sample include NGC 4569, which 
has a very blue nucleus (see also Keel 1996), NGC 4694, and NGC 4527. 

Ideally, foreground unsaturated stars should disappear if 
the images have been carefully registered and the seeing matched.
However, the point spread 
functions are often non-Gaussian and different in the R
and H$\alpha$ filters due to a number of factors, including
the difference in the thicknesses of the R and H$\alpha$ filters,
slight collimation errors, slight guiding errors, and differences in
the focus between the two frames. Therefore, residuals often 
remain on the brighter stars due to imperfect continuum subtraction.
(In some cases, there are systematic patterns in the stellar residuals which
can be used as a diagnostic for the accuracy of the telescope collimation.)
The relative fluxes of the star in the R image and the residuals
in the H$\alpha$ continuum subtracted image (summing the absolute value
of counts) are usually on the order of 1-2\%.  
Residuals near the galaxy were removed using the pixel replacement routines 
in the IRAF task \it imedit \rm.

Examination of the stellar residual pattern for images taken during the
February 1995 run at KPNO 0.9-m telescope revealed a slight mismatch 
between the pixel scales of the R and H$\alpha$ images.
The IRAF tasks geomap and geotran were used to rescale the images
to the same pixel scale before continuum subtraction.
 
The use of a broadband R filter as a substitute for an off-line narrowband 
filter allows significant observing time savings, making a survey of this 
size practical.
However, there is a trade-off in accuracy (see Pogge 1992) for two reasons.
Broadband and interference filters often differ in thickness and response
to light, so can cause differing point-spread functions and even
small plate scale differences, as noted above.
In addition, the larger
bandwidth of the R filter means that the derived value of the continuum is less
certain and may be subject to color effects across the filter. Even small
uncertainties in the continuum level are important, since they cause
large uncertainties
in the H$\alpha$ flux, especially in regions of weak or diffuse emission.
To estimate the extent of error due to the use of a broadband filter rather
than a narrowband filter, we compared 2 H$\alpha$ continuum subtracted 
images of the galaxy NGC 4102, kindly provided to us by Shardha Jogee. These
images were taken during our February 1995 run.
In one, the continuum was derived from an image taken with a narrowband 
H$\alpha$ offline filter ($\lambda$ = 6653 \AA, $\Delta \lambda$ = 68 \AA) 
and in the 
other, from a Harris R image. Stellar residuals are present in both 
continuum-subtracted images and differ by 2-3\% in the total flux 
(summing the absolute value of residual counts).
The total H$\alpha$ flux of the galaxy is the same to within 3\%. The radial
profiles in H$\alpha$ have the same shape and no systematic dependence on
the continuum-subtracting filter. Aperture photometry of low H$\alpha$
surface brightness areas in the galaxy results in a 5-10\% difference between
the images, with the offline filter producing lower fluxes.
A significant
part of these differences can be attributed to uncertainties in the sky level
and scale factor in each image. Based on these 
tests, we conclude that the R filter does a reasonable job at 
continuum subtracting H$\alpha$, with error contributions smaller than 
the other sources of uncertainty, with the probable exception of galaxies
with very weak H$\alpha$ emission. 

In summary, systematic errors typically cause a 20-30\%
uncertainty in total H$\alpha$ fluxes and surface brightnesses.

\subsection{Stellar Masks}
Before surface photometry can be performed on an R image, the stars must
be removed or masked. DAOPHOT (Stetson 1987) was used to find stars
across the image. 
Rather than removing the stars (which involves an iterative determination of 
the point spread function across the chip and does not ever fully remove
stars), we chose in most cases
to make a pixel mask image. In the first pass, sources found 
within the area of the galaxy disk were excluded from the mask, 
since HII regions are also detected by DAOPHOT.
Stars superposed on the disk and stars missed by DAOPHOT were masked by hand
using IRAF centering and the \it badpiximage \rm routine. Members of galaxy
pairs were masked in a similar manner before surface photometry was derived.
Mask images were also necessary for
the H$\alpha$ image in the case of galaxy pairs or severe stellar 
residuals from stars which were saturated in the R image.

\section{The Images}
\label{images}
R and H$\alpha$ images and surface photometry are presented for each galaxy in 
Figure~\ref{virim}.
Galaxies are ordered according to NGC/IC number, with morphological 
information indicated on the plots. Derived photometric scalelengths are
indicated in the surface photometry plots, as described in the figure
caption. 

\section{Surface Photometry}
\label{surfphot}
Deriving a
radial profile of a galaxy consisted of several steps: determination of
center, the axial ratio (ratio of minor to major axis lengths) and 
position angle, careful measurement of 
sky background error, and the surface photometry.
In this section, we discuss the derivation of the
radial profiles, quantities derived from the profiles, and the errors. 

\subsection{Determining Centers, Axial Ratios, and Position Angles}
\label{sppar}
The center of the
galaxy was determined from the R image using centroiding via the
\it center \rm task in IRAF. 
Several of the galaxies (NGC 4299, NGC 4419, NGC 4424) had
no obvious central peak or had a central region 
contaminated by bright HII regions.
In these cases, the center was determined from
inner isophotes or midpoints between peaks. 

Radial profiles were derived by assuming fixed values of the axial ratio and 
position angle, which were determined by examination of the outer isophotes. 
Starting values
of axial ratio and position angle were determined by eye from the outer 
isophotes of the galaxy 
and/or literature inclination values 
(e.g., RC3, Warmels 1988). 
For several galaxies, including 2 of low inclination, axial ratios and/or
position angles determined from HI maps (Warmels 1988) were used. 
For other galaxies, the
axial ratios and position angles were
adjusted as necessary to give a good fit to the outer isophotes. 

The conversion of axial ratios to inclination angles is not 
well-determined (e.g. Peletier \&Willner 1991). The standard conversion is
$cos^2 \ i =  (q^2 - q_0^2)/(1-q_0^2)$, where
i is the inclination, q is the axial ratio, and q$_0$ is the
intrinsic axial ratio (Hubble 1926). 
The intrinsic axial ratio is often assumed to be 
0.2 (Holmberg 1958) but it may vary with type (e.g. Bottinelli et al. 1983). 
The uncertainty in the conversion to inclination is most
serious for galaxies of high inclination and so is not a major source
of error for this study, since most sample galaxies have 
inclinations less than 75$^{\circ}$.
For example, the difference in inclination calculated by assuming an
intrinsic axial ratio of 0 versus 0.2 is 3$^{\circ}$ at 70$^{\circ}$. 
(In Section~\ref{errors},
we show the uncertainty in the radial surface brightness profiles 
is minor for a 5$^{\circ}$ error in the inclination.)

Adopted axial ratios and
position angles are listed in columns (2) and (3) of Table~\ref{rtab}.
The inclination calcalated using the standard conversion with
intrinsic axial ratio of 0.2 is given in parentheses after the axial ratio.
The random uncertainties in the estimation of the inclination and position 
angle from the outer isophotes are typically 2-3 $^{\circ}$, except in the 
indicated cases.
Galaxies which have derived inclination and position angle significantly
different from literature values are noted in the table. 
The H$\alpha$ profile
was measured using the same center, inclination, and position angle as for 
the R image. 

\placetable{rtab}

\begin{figure}
\centerline{\includegraphics[height=4in]{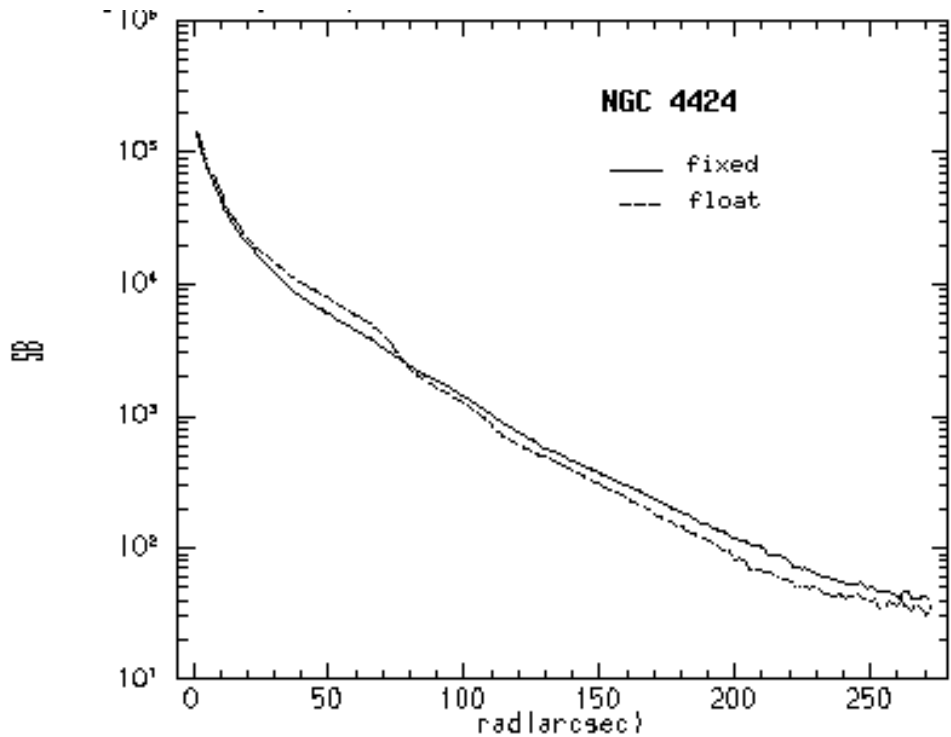}}
\caption
{R Radial profiles of the galaxy NGC 4424. Surface brightness is
in units of 10$^{-18}$ erg cm$^{-2}$ s$^{-1}$. The solid profile is
determined using fixed center, inclination (54), and position angle (90), while
the dashed profile was derived allowing center, inclination, and position
angle to `float' within 130$^{\prime\prime}$. The difference in the profiles
is small and does not significantly affect comparisons between large 
numbers of galaxies, nor the derivation of central light concentrations. 
\label{fixfl}}
\end{figure}

In most cases, R profiles of the galaxies were also derived by allowing
the position angle, axial ratio, and center to vary with radius. This
approach is useful in studying the variation in the position angle of bars, 
rings,
and the intrinsic variation in the position angle or axial ratio, 
which might indicate a recent
dynamical disturbance. Position angle and/or axial ratio differences with
radius are present in several galaxies in our sample.
However, elliptical annuli which are allowed
to vary with radius are influenced by 
spiral structure and are prone to errors in low surface
brightness regions. In addition,
the intention of this study is not to
study galaxy structure in detail, but to
obtain a general profile for each galaxy, which will allow us to 
compare the profiles and derived quantities such as concentration. For our
purposes, the differences between profiles derived using fixed elliptical
annuli and those using varying elliptical annuli are mostly small. A 
typical example is shown in Figure~\ref{fixfl}.

\subsection{Derivation of Profiles}
With the center, axial ratio, position angle, and sky background uncertainties
determined (as described in Section~\ref{skyerr}), radial profiles were 
derived using a surface photometry
program written within the Interactive Data Language environment, 
hereafter \it SPHOT. \rm
\it SPHOT \rm allows determination of realistic
errors, input of a bad pixel mask image, 
a correction for masked pixels to derive total flux, and simultaneous
measurements of isophotal radii, concentration parameters, and integrated 
fluxes.

The center, axial ratio, position angle, and sky level uncertainty
are read into \it SPHOT \rm.
If there is a mask image, the input image is first multiplied by the mask.
Surface photometry is begun at a radius equal to 
one-half of the FWHM. The minimum annulus width is
equal to the FWHM, and the width increases by 10\% of the FWHM 
for each annulus, so that larger annuli are used in the outer regions where
the S/N is smaller.
The program calculates the surface brightness based on the unmasked pixels 
whose centers fall within a
given annulus. The area of the 
annulus is calculated using the semimajor
and semiminor axes of the annulus and the flux is calculated by multiplying 
the surface brightness by this area. This compensates for the flux from 
fractional and masked pixels within the annulus. 

The R radial profiles were calculated out to a radius where 
the error in the surface brightness exceeded the signal. The errors
in radial profiles are discussed in the next section. 

The outermost
radius of visible HII regions was determined in advance from inspection of the
images, and the H$\alpha$ profile was halted just outside this radius. 
In some cases,
especially for galaxies with faint outer disk star formation or galaxies
with scattered background light or poor flat fields, this
radius was beyond the point in the profile where sky uncertainty errors
overwhelmed the azimuthally averaged signal. In the plots of radial profiles
in Figure~\ref{virim}, we indicate this
radius, but plot the H$\alpha$ radial profile to the outermost HII region
to emphasize the extent of star formation. 

All profiles were corrected to face-on assuming complete transparency in the
disk, i.e., the correction 2.5log(a/b) was applied. The assumption of complete
transparency is oversimplified, especially at inner radii, but since
correction factors for internal
extinction are poorly known, no corrections were made.
(See Giovanelli et al. 1994 for more discussion.) 
 
The inclination corrected radial profiles of individual galaxies are plotted 
with the images of the galaxies in Figure~\ref{virim}.

\subsection{Errors in the Radial Profiles}
\label{errors}
There are several sources of errors in the calculation of the surface 
brightness at each radius. Processing errors
contribute a relatively small percentage of the total error, and the
readnoise of the chips is low. 
Random error is the dominant source of error in the inner
regions of the surface brightness profiles and an important source of error 
in the integrated flux measurements. 
It is traced by the standard deviation in the background (e.g.,
Newberry 1991), which was determined using \it fitsky
\rm or \it imstat \rm in IRAF. In the outer parts of the profiles,
large numbers
of pixels are averaged to give the surface brightness, so the random error
is small. In the ideal case of an accurately determined sky, the total error 
would be due only to the random error and would be equal to 
the standard deviation of the mean. However, the
systematic error in determining the sky brightness overwhelms the small 
random error in the outer regions. It is therefore critical to obtain as
flat a background as possible for deep surface photometry. The total error
in the profiles was calculated by adding in quadrature the contributions from
the random error and the sky uncertainty (Section~\ref{skyerr}). 

\begin{figure}
\centerline{\includegraphics[height=7in]{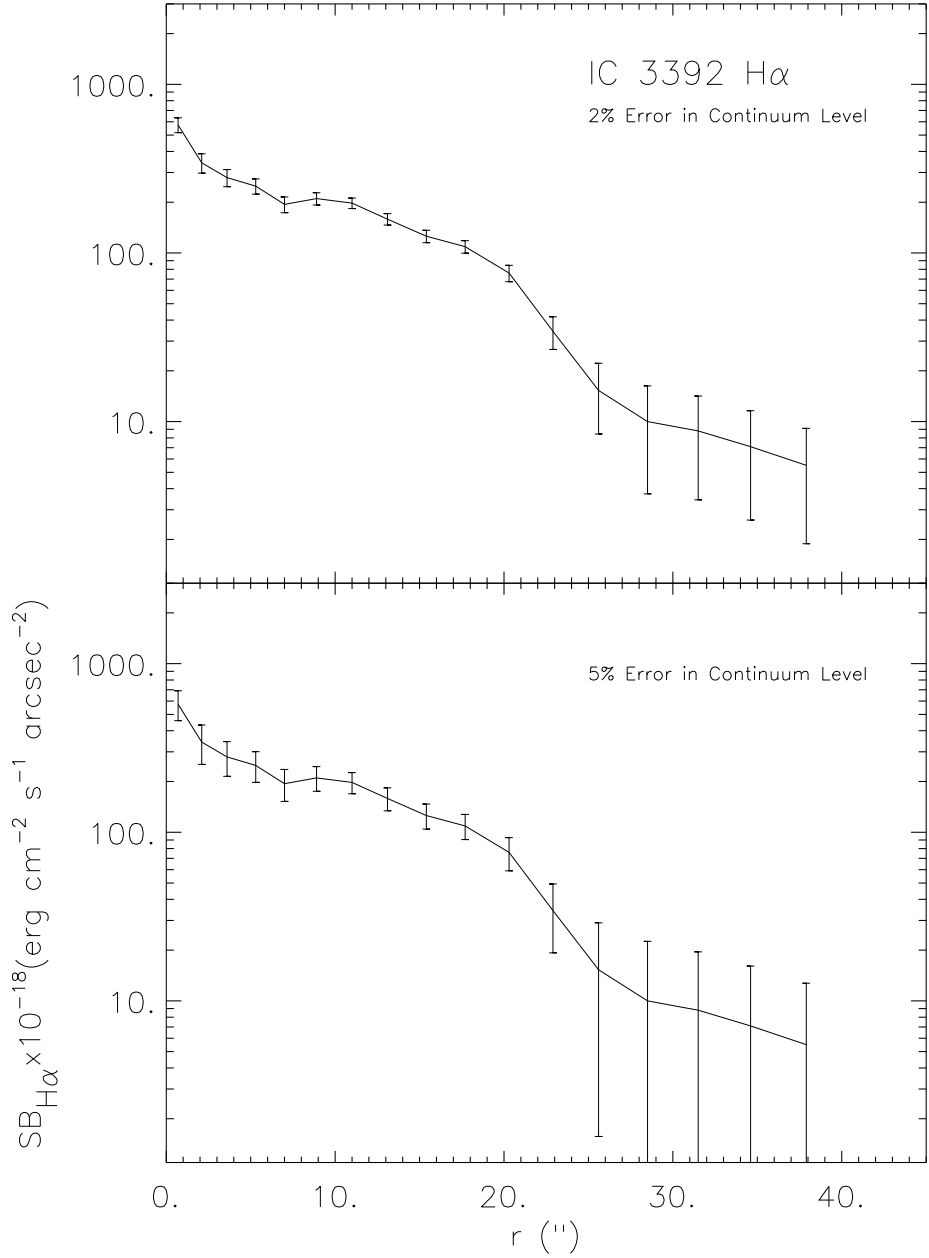}}
\caption
{The effect of a 2\% (upper) and 5\% uncertainty in the continuum subtraction 
level on the H$\alpha$ radial profiles of IC 3392. Even small uncertainties
in the continuum subtraction value can result in large uncertainties in the
H$\alpha$ flux and surface brightness. \label{radercs}}
\end{figure}

Not included in the error calculation for individual galaxies are errors in
the absolute flux calibration ($\sim$ 5\%) and errors in the continuum 
subtraction. Figure~\ref{radercs} shows typical errors in the H$\alpha$
radial profile due to 2\% and 5\% errors in the continuum level. This
Figure illustrates the sensitivity of H$\alpha$ surface brightness to
even small uncertainties in the continuum scale factor.

\begin{figure}
\plotone{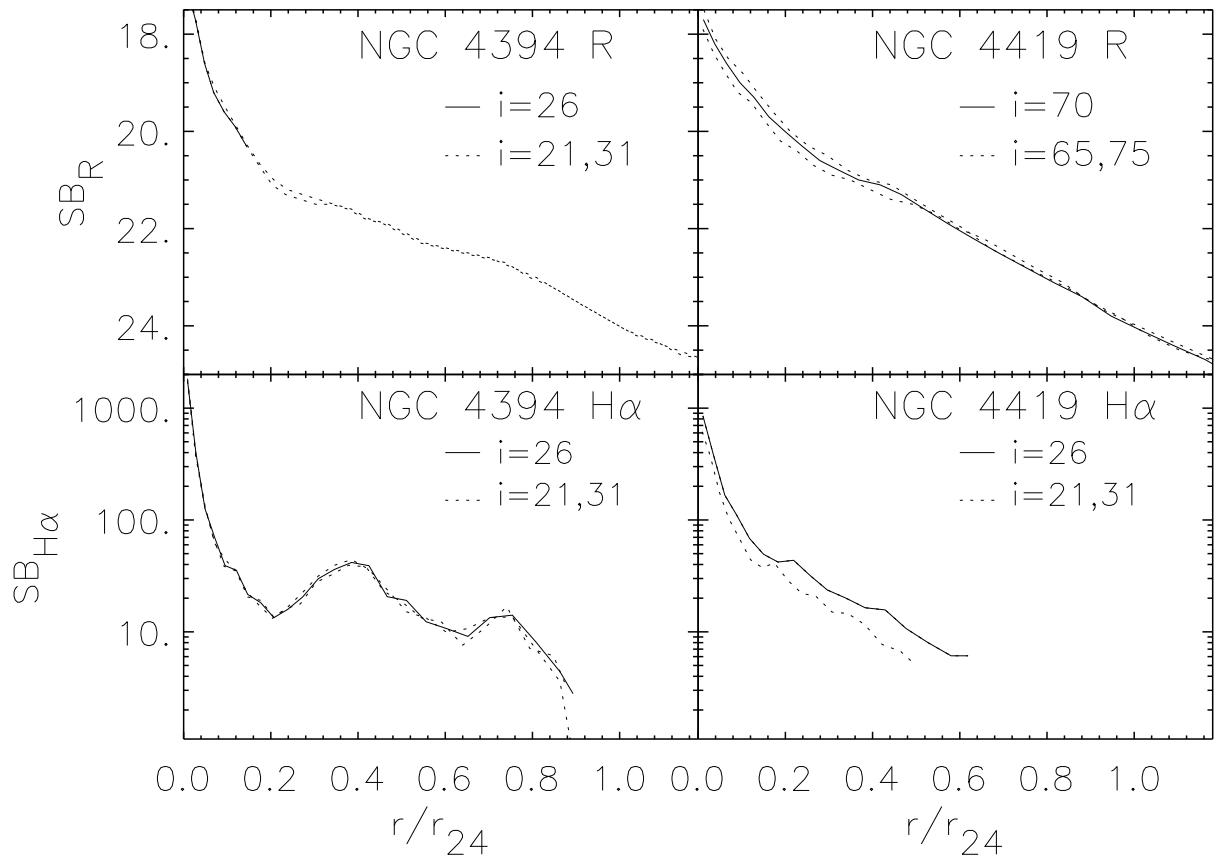}
\caption
{The effect of inclination error of 5 degrees on R and H$\alpha$ 
radial profiles of NGC 4394 (left), which has an 
inclination of 26$^{\circ}$, and NGC 4419 (right), which has an 
inclination of 70$^{\circ}$.
Uncertainties in the radial profile due to inclination uncertainty 
are most severe for highly inclined galaxies. However inclination 
uncertainties do not strongly affect comparisons of profiles between
galaxies nor derivation of central concentration.
\label{incer}}
\end{figure}

The radial profile is sensitive to errors in the inclination. Inclination
errors are largest for 
highly inclined galaxies and galaxies with strong
spiral arms. For most galaxies in this sample, the inclination is uncertain
to less than 3$^{\circ}$, as determined from the outer isophotes. The typical
variation in the radial profile due to an error of 5$^{\circ}$ in inclination
is shown for a high inclination (70$^{\circ}$) galaxy, NGC 4419, and a 
low inclination (26$^{\circ}$) galaxy, NGC 4394 in Figure~\ref{incer}. 
The uncertainty in the radial
profile is small even in the case of the high inclination galaxy. 

\subsection{Derived Quantities}
\label{derq}

\subsubsection{Isophotal Radii and Integrated Flux Measurements} 
A quantity useful as a galaxy size indicator is an outer isophotal radius
measured in a broadband filter. R isophotal radii at 24 and, if possible, at 
25
mag arcsec$^{-2}$, were determined by interpolation of the radial profiles. 
90\% of the galaxies have measured radii at the 25 mag arcsec$^{-2}$
isophote. Table~\ref{rtab} 
lists the isophotal radii in units of arcsecs in columns
(4) and (6) and the corresponding total R fluxes within the isophotal radius
in columns (5) and (7). 

Total H$\alpha$ fluxes were measured in three different ways. 
Total fluxes to the radius of the outermost HII region were
measured through a polygonal
aperture, defined by the extent of H$\alpha$ emission, using the \it polyphot
\rm task in IRAF. This result can be compared to the total flux measured within
the outermost elliptical annulus, measured within the surface photometry
program. Because the sensitivity varied between images,
we also measured an integrated flux within the radius, r$_{H\alpha17}$,
at which the surface brightness 
falls to 17x10$^{-18}$ erg cm$^{-2}$ s$^{-1}$, a sensitivity level
met by all the images. The three measures of the H$\alpha$ flux are comparable.
The total flux determined by surface photometry and aperture photometry
are the same within a few percent. The total flux is on the average
10\% higher than the flux within r$_{H\alpha17}$, but with a scatter (standard 
deviation) among galaxies of about 16\%. This is similar to the isolated 
sample,
which has total fluxes about 20\% higher than the flux within r$_{H\alpha17}$,
with a 20\% scatter. Two Virgo galaxies
with low surface brightness, NGC 4643 and 4698, had total fluxes
3 and 9 times, respectively, higher than the flux to r$_{H\alpha17}$.
Finally, the H$\alpha$ flux within 0.3r$_{24}$ and H$\alpha$ concentration
(Section~\ref{haconc}) were measured.
Table~\ref{hatab} presents quantities derived from the 
H$\alpha$ surface photometry. 

\placetable{hatab}

\subsubsection{Disk Scalelengths}
\label{diskscale}
An isophotal radius is one type of radial normalizer. Another radial normalizer
which suffers less from intrinsic surface brightness variations between
galaxies is the exponential disk scalelength. 
The derivation of disk scalelengths is
less subject to sky background problems and small frame size than derivation
of isophotal radii, since the whole radial range of the light profile is used.
However, the measurement of a disk scalelength
is more complicated than that of a simple isophotal
radius, because it is usually derived by non-linear fitting of an exponential
function to the radial profile. 
Fitting bulge and disk models to radial profiles
is a tricky task, subject to the form of the models chosen, uncertainties
due to the projection of a three-dimensional distribution into the sky,
seeing corrections,
contamination of the profiles by star forming regions, bars, rings, dust
(e.g., Giovanelli et al. 1994, 1995), and
uncertainties in the background values. The greatest uncertainties are
in fitting a proper bulge, especially given the discussion over exponential
vs. deVaucouleurs laws (see Kormendy 1992, Courteau et al. 1996, 
de Jong 1996a). 
In order to derive a disk scalelength,
we fit traditional r$^{\frac{1}{4}}$ bulge and exponential
disk models to the profiles, using the IRAF/STSDAS non-linear fitting task,
\it nfit1d \rm.
We warn that extreme care must be used in applying this task,
since it is very sensitive to complexities in the radial profile.
The task attempts
to compensate for deviations from a pure bulge + disk profile
by overestimating the contribution of the bulge and/or 
by overemphasizing the inner disk fit compared to the outer disk fit. 
We report only the disk scalelength because of the
4 parameters that varied in the bulge-disk decompositions, 
we find that this parameter is
the most stable. (This was also noted by de Jong 1996b, who find that r$_d$
is more stable than the central disk surface brightness.) Even so, 
different authors often derive different
scalelengths, so we attempted to test the full range of possible disk
scalelengths by fitting different radial ranges with different assumptions
for the central component. Among isolated galaxies (Koopmann \& Kenney 2001a)
and most of the Virgo
Cluster sample,  the disk scalelength typically varies by 10-15\%
for the different fits. The central disk surface brightnesses are more
dependent on the assumed form of the central component. 
The disk scalelength and estimate of uncertainty in arcseconds
appear in columns (10) and (11) of Table \ref{hatab}. 
Disk scalelengths could
not be fit to two Virgo Cluster galaxies (NGC 4383, NGC 4586). The
Virgo sample has a higher proportion of complex profiles:
20\% (11/53) of Virgo
Cluster galaxies have derived disk scalelengths which are uncertain by
$>$ 20\%, compared to 0\% (0/27) of the isolated sample galaxies 
(Koopmann \& Kenney 2001a).   
 
If the surface brightness of disks were self-similar and all disks were
exponential with measurable scalelengths, r$_{24}$ would be perfectly
correlated with r$_d$. We find that the correlation between these two
quantities is good, with a measured ratio of 
r$_{24}$/r$_d$ of
3.23$\pm$0.72 for the Virgo Cluster, compared to
3.16 $\pm$ 0.56 for the sample of isolated galaxies (Koopmann
\& Kenney 2001a). These values are similar to that found by 
Giovanelli et al. (1995).

Given the systematic effects
of differing surface brightnesses 
on isophotal radii and the difficulties of model fitting
and unknown extinction values on disk scalelength, it is not obvious which
radial normalizer is a better indicator of galaxy/disk size. 
We compared results based on both normalizers, but 
found that our main conclusions are independent of the normalizer. 
Because more galaxies could be included in the analysis and because of
the more uncertain nature of disk scalelength fitting,
we chose to normalize by r$_{24}$.

\subsubsection{R Concentration Parameter}
\label{conc}
Because of the subjectivity of the Hubble classification and 
the difficulty in  
deriving model-independent bulge-to-disk (B/D) ratios 
(section~\ref{diskscale}),
we also measured a central R light concentration index.

Central light concentration parameters are objective tracers of
the radial light distribution of a galaxy (e.g., Morgan 1958; 
deVaucouleurs 1977; 
Okamura et 1984; Kent 1985; Abraham et al. 1994, Koopmann \& Kenney 1998).
Concentration parameters are correlated to B/D or 
bulge-to-total light ratio (deVaucouleurs 1977; Okamura et 1984; 
Kent 1985; Eder 1990, but see Naim et al. 1997) and Hubble type 
(Abraham et al. 1994) of spiral galaxies, although they may be be less 
sensitive in differentiating Sa, S0,
and E galaxies (Abraham et al. 1994, Smail et al. 1997, van den Bergh 1997)
and they are somewhat dependent on a galaxy's surface brightness (e.g.,
Abraham et al. 1994). 
Unlike B/D, concentration parameters are
independent of models assumed for galaxy components. 

We define a central R light
concentration parameter similar to Abraham et al. (1994):
$$\rm C30=\frac{F_R(0.3r_{24})}{F_R(r_{24})}$$
where F$_R$(r$_{24}$) is the flux in R measured 
within the r$_{24}$
isophote and F$_R$(0.3r$_{24}$) is the flux within the 0.3r$_{24}$ isophote.
The C30 values are computed directly from the surface photometry.
We find no correlations between C30 and total
R magnitude within r$_{24}$, inclination of the galaxy, or
angular distance from M87. There is a mild dependence on surface 
brightness, which 
causes higher surface brightness galaxies to have a higher C30. 

C30 can be approximately translated to B/D using
the following mean [C30,B/D] pairs: [0.3,0.1], [0.4,0.3], [0.5,0.8], 
[0.6,1.0]. These numbers are meant to only suggest a rough scale for C30,
particularly since
the scatter of these numbers is at least $\pm$0.1, due to uncertainties
in the determination of both C30 and B/D.
The correlation between central light concentration and Hubble type 
is discussed in Koopmann \& Kenney (1998, 2001b).

\subsubsection{H$\alpha$ Concentration Parameter}
\label{haconc}

In order to compare the concentrations of R and H$\alpha$ light,
we measure a central H$\alpha$ light concentration analogous to C30:

\begin{displaymath}
\rm CH\alpha = \frac{F_{H\alpha}(0.3r_{24})}{F_{H\alpha}},
\end{displaymath}

\noindent where F$_{H\alpha}$ is the total H$\alpha$ flux 
and F$_{H\alpha}$(0.3r$_{24}$) is the flux within the 0.3r$_{24}$ isophote.
This quantity is listed in Column 7 of Table~\ref{hatab}. 
A CH$\alpha$ of 1 thus indicates that all of the H$\alpha$ emission is
located within 0.3r$_{24}$. 

To compare the relative distributions of R and H$\alpha$ light, we
normalize CH$\alpha$ by C30: $\frac{CH\alpha}{C30}$.
For example, this quantity is larger than 1 when the H$\alpha$ emission is more
concentrated than the R light. 
This quantity is listed in Column 8 of Table~\ref{hatab}. 
H$\alpha$ concentrations are compared
for the two samples in Koopmann \& Kenney (2001b).

\section{Discussion}
Further quantitative analyses of the spatial distributions of massive
star formation in Virgo Cluster and isolated spirals are provided in
Koopmann \& Kenney (1998, 2001a, 2001b). Here we comment briefly on
the unusual H$\alpha$ morphology of several spirals in the Virgo Cluster
sample.

It is evident from our H$\alpha$ images (Figure~\ref{virim})
that there is a population of
galaxies in the Virgo cluster with active star formation in the
circumnuclear regions, but no other disk star formation. NGC 4064,
NGC 4351, NGC 4405, NGC 4424, NGC 4522, NGC 4580, and IC3392 are 
examples of this class.
Koopmann \& Kenney (2001b) quantitatively define this class as low to
intermediate light concentration 
galaxies which have star formation rates which are similar to isolated
spirals in the circumnuclear
regions, but which have no star formation in the outer 60-70\% of the optical
disk. We call these galaxies St (spirals with severely truncated 
star-forming disks), since they do not fit the standard Hubble classification
criteria (see Koopmann \& Kenney 1998).
 
St galaxies are extreme examples of what 
van den Bergh et al. (1990) call 'Virgo-type' galaxies, galaxies with
'with fuzzy outer regions
which exhibit active star formation in their central bulges (disks?)'.
The definition of this class was rather broad, including
NGC 4351, NGC 4424, and IC 3392, but also galaxies such as NGC 4212, NGC 4689,
and NGC 4527,
which have more extended star formation in the outer disk 
(see Figure~\ref{virim}). 
The more general 'Virgo-type' morphology is due to the truncation of
the star-forming disk in cluster galaxies, which is the primary cause
of the reduction in star formation rates (Kennicutt 1983) in Virgo 
Cluster galaxies (Koopmann \& Kenney 2001b). 

The extreme truncation of St star-forming disks, coupled with the active
inner star formation is an indication that these galaxies have had
interesting evolutionary histories. NGC 4424 shows evidence of a recent
merger event (Kenney et al. 1996). NGC 4522 is currently being stripped by
the intracluster medium (Kenney \& Koopmann 1999). 
Others among the St class may be remnants of
past intracluster medium stripping events.

\acknowledgements
The funding for this research was provided by NSF grant AST-9322779.
We are grateful to Con Deliyannis for his help in obtaining observations
of standard stars for the derivation of the transformation equation and
Sydney Barnes, Y.-C. Kim, Raj Jain, 
and Charles Bailyn for obtaining several of the images in this paper during
service observing runs. We thank R. Pogge for excellent advice on obtaining 
the best continuum subtractions and R. Kennicutt for advice and encouragement.
Comments from an anonymous referee led
to several improvements in the final draft.  
This work was substantially aided by
observing support from the Kitt Peak and Cerro Tololo staff.
This research has made use of the NASA/IPAC Extragalactic Database (NED)
which is operated by the Jet Propulsion Laboratory, California Institute
of Technology, under contract with the National Aeronautics and Space
Administration. The authors and maintainers of the IRAF and STSDAS software
packages and instruction manuals are also gratefully acknowledged.

\onecolumn

\newpage

\begin{figure}
\caption
{The R and H$\alpha$ images and surface photometry.
Galaxies are ordered according to NGC/IC number. Images for which
radial profiles were not derived appear at the end of the sequence.
The images are displayed on a log scale, with north up and east to the
left. The tickmarks on the images are spaced by 1.0 arcmin, and the solid line
on each image represents 1 arcmin, which is equivalent to about 5 kpc
for a Virgo Cluster distance of 16 Mpc. Sky around some of the
larger field-of-view images was cropped; refer to Tables 3 and 4 for
the actual size of the image frames.
The RSA/BST morphology class is noted in the righthand corner of each image
and the RSA/BST and RC3 morphological types are indicated in the surface
photometry plots. 
In the surface photometry plots, the R (solid) and 
H$\alpha$ (dotted) profiles are plotted as a function of radius in arc seconds.
The H$\alpha$ profiles were superposed using an arbitrary
zeropoint of 18.945. A solid line indicating 1 kpc is provided below the 
morphological types. 
The isophotal radius at 24 mag arcsec$^{-2}$, r$_{24}$ 
and the disk scalelength, r$_d$, are indicated with arrows. An error bar
for the R profile is given at r$_{24}$, and the R profile ends where the
noise becomes greater than the signal. The H$\alpha$ profile is cut at
the radius of the outermost HII region. Diamonds on the H$\alpha$
profile indicate annuli for which the sky uncertainty was greater than the 
azimuthally averaged signal. In addition, a circle is plotted on the
H$\alpha$ profile 
at the 17 x 10$^{-18}$ erg cm$^{-2}$ s$^{-1}$ arcsec$^{-2}$
isophotal radius (which corresponds to a value of 23 R mag arcsec$^{-2}$ in
the offset scale. \label{virim}} 
\end{figure}

\begin{deluxetable}{lllcclrrr}
\tablecaption{Properties of Observed Virgo Cluster Galaxies}
\tablewidth{0pt}
\tablehead{
\colhead{(1)}& \colhead{(2)}&\colhead{(3)} &\colhead{(4)} &\colhead{(5)} &\colhead{(6)} & \colhead{(7)} & \colhead{(8)} & \colhead{(9)}\\
\colhead{Name}& 
\colhead{RA (1950)}& 
\colhead{Dec (1950)}& 
\colhead{RSA/BST}&
\colhead{RC3}&
\colhead{B$_T^O$}&
\colhead{v$_{he}$}&
\colhead{D$_{87}$}&
\colhead{HI }\\
\colhead{}&
\colhead{(h m s)}&
\colhead{(d m s)}&
\colhead{}&
\colhead{}&
\colhead{}&
\colhead{(km/s)}&
\colhead{($^{\circ}$)}&
\colhead{Def}}
\startdata
NGC 4064&12 01 37.8&18 43 16&SBc(s):&SB(s)a:pec&12.30&913&8.8&.99\\
NGC 4178&12 10 13.2&11 08 38&SBc(s)II&SB(rs)dm&11.89&378&4.7&-.13\\
NGC 4189&12 11 14.5&13 42 11&SBc(sr)II&SAB(rs)cd&12.53&2115&4.3&.20\\
NGC 4192&12 11 15.5&15 10 42&SbII:&SABab&10.92&-142&4.8&.19\\
NGC 4206&12 12 44.0&13 18 07&Sc(s)&Sbc:&12.77&702&3.8&.09\\
NGC 4212&12 13 06.6&14 10 46&Sc(s)II-III&Sc:&11.86&-81&4.0&.44\\
NGC 4216&12 13 21.5&13 25 40&Sb(s)&SABb:&10.97&131&3.7&.55\\
NGC 4237&12 14 38.9&15 36 07&Sc(r)II.2&SAB(rs)bc&12.37&867&4.4&.62\\
NGC 4254&12 16 17.0&14 41 39&Sc(s)I.3&Sc&10.43&2407&3.6&.02\\
NGC 4293&12 18 41.0&18 39 35&Sa pec& (R)SB(s)0/a pec&11.20&893&6.4& $>$ 1\\
NGC 4294&12 18 45.3&11 47 10&SBc(s)II-III&SBcd&12.62&355&2.5&-.08\\
NGC 4298&12 19 00.6&14 53 01&Sc(s)III&Sc(rs)&12.08&1135&3.2&.54\\
NGC 4299&12 19 08.0&11 46 48&Scd(s)III&SABdm:&12.86&232&2.4&.05\\
NGC 4302&12 19 10.1&14 52 30&Sc(edge)&Sc:&12.55&1149&3.1 &.43\\
NGC 4303&12 19 21.6&04 45 03&Sc(s)I.2&SAB(rs)bc&10.17&1566&4.8&.17\\
NGC 4321&12 20 22.9&16 05 58&Sc(s)I&SABbc&10.11&1571&3.9&.52\\
NGC 4351&12 21 29.1&12 28 54&Sc(s)II.3&SB(rs)ab:pec&13.04&2310&1.7&.58\\
NGC 4380&12 22 49.7&10 17 38&Sab(s)&Sb(rs):&12.36&967&2.7&1.1\\
NGC 4383&12 22 53.8&16 44 49&Amorph&Sa pec&12.68&1710&4.3&-.39\\
NGC 4394&12 23 24.3&18 29 26&SBb(sr)I-II&RSBb?&11.76&922&5.9&1.1\\
NGC 4405&12 23 35.5&16 27 28&Sc/S0&SA(rs)0/a&12.99&1747&4.0&1.1\\
NGC 4411B&12 24 14.8&09 09 41&Sc(s)II&SABcd&12.92&1270&3.6&.31\\
NGC 4413&12 24 00.2&12 53 16&SBbc(rs)II-III&SBab:&12.97&102&1.1&.90\\
NGC 4419&12 24 24.7&15 19 24&Sa&SBa(sp)&12.13&-261&2.8&1.1\\
NGC 4424&12 24 39.0&09 41 51&Sa pec&SAB0p&12.32&439&3.1&1.1\\
NGC 4429&12 24 54.1&11 23 05&S0$_3$(6)/Sa pec&SA(r)0$^+$&10.9&1130&1.5&$>$ 1\\
NGC 4438&12 25 13.6&13 17 07&Sb(tides)&SA(S)0/a pec&10.91&71&1.0&1.6\\
NGC 4450&12 25 58.0&17 21 42&Sab pec& SAab&10.93&1954&4.7&1.3\\
 IC 3392&12 26 11.7&15 16 34&Sc/Sa&SAb:&13.30&1687&2.7&1.4\\
NGC 4457&12 26 26.0&03 50 51&RSb(rs)II&(R)SAB(s)0/a&11.76&882&8.8&.83\\
NGC 4459&12 26 28.3&14 15 20&S0$_3$(2)&SA0(late)&11.37&1210&1.6&$>$ 1\\
NGC 4498&12 29 08.6&17 07 41&SBc(s)II&SABd&12.62&1507&4.5&.37\\
NGC 4501&12 29 27.5&14 41 43&Sbc(s)II&SAb&10.27&2281&2.0&.47\\
NGC 4519&12 30 58.0&08 55 49&SBc(rs)II.2&SBd&12.34&1220&3.8&-.28\\
NGC 4522&12 31 07.6&09 27 03&Sc/Sb:&SBcd:(sp)&12.73&2328&3.3&.60\\
NGC 4526&12 31 30.4&07 58 33&S0$_3$(6)&SAB0:&10.61&448&4.8&$>$ 1\\
NGC 4527&12 31 35.5&02 55 45&SB(s)II&SABbc&11.32&1736&9.8&-.35\\
NGC 4532&12 31 46.7&06 44 39&SmIII&Im&12.30&2012&6.0&-.30\\
NGC 4535&12 31 48.0&08 28 26&SBc(s)I.3&SABc&10.51&1961&4.3&.17\\
NGC 4536&12 31 53.5&02 27 50&Sc(s)I&SABbc&11.01&1804&10.2&.03\\
NGC 4548&12 32 55.2&14 46 20&SBb(rs)I-II&SBb&10.98&486&2.4&.86\\
NGC 4561&12 33 38.0&19 35 53&SBcIV&SBdm&12.96:&1407&7.0&-.33\\
NGC 4567&12 34 01.0&11 31 59&Sc(s)II-III&SAbc&12.08&2274&1.8&.64\\
NGC 4568&12 34 02.5&11 30 50&Sc(s)III&SAbc&11.70&2255&1.8&.64\\
NGC 4569&12 34 18.5&13 26 16&Sab(s)I-II&SABab&10.25&-235&1.7&.99\\
NGC 4571&12 34 25.3&14 29 33&Sc(s)II-III&SAd&11.81&342&2.4&.44\\
NGC 4579&12 35 12.0&12 05 34& Sab(s)II&SABb&10.56&1519&1.8&1.0\\
NGC 4580&12 35 15.9&05 38 36&Sc/Sa&SABa?&12.49&1034&2.5&1.3\\
NGC 4586&12 35 55.1&04 35 37&Sa&SAa:&12.54&794&8.3&1.2\\
NGC 4606&12 38 26.2&12 11 09&Sa pec&SBa:&12.69&1664&2.5&$>$1\\
NGC 4607&12 38 41.0&12 09 36&...&SBb?sp&12.79&2257&2.6&.74\\
NGC 4639&12 40 21.5&13 31 52&SBb(r)II&SABbc&12.19&1010&3.1&.16\\
NGC 4643&12 40 46.9&02 15 06&...&SB(rs)0/a&11.54&1399&10.9&$>$ 1\\
NGC 4647&12 41 01.1&11 51 20&Sc(rs)III&SABc&12.03&1422&3.2&.51\\
NGC 4649&12 41 08.4&11 49 34&S0$_1$(2)&E2&9.81&1413&3.2&$>$ 1\\
NGC 4651&12 41 12.4&16 40 01&Sc(r)I-II&SAc&11.36&805&5.1&-.16\\
NGC 4654&12 41 25.9&13 23 59&SBc(rs)II&SABcd&11.14&1035&3.3&0\\
NGC 4689&12 45 15.0&14 02 10&Sc(s)II.3&SAbc&11.55&1616&3.7&1.1\\
NGC 4694&12 45 44.0&11 15 28&Amorph&SB0 pec&12.19&1175&4.5&1.2\\
NGC 4698&12 45 51.3&08 45 35&Sa&SAab&11.53&1002&5.8&.25\\
NGC 4710&12 47 09.0&15 26 15&S0$_3$(9)&S0$^+$?(sp)&11.85&1129&5.4&$>$ 1\\
NGC 4713&12 47 25.4&05 34 59&SBc(s)II-III&SABd&12.21&653&8.5&-.13\\
NGC 4772&12 50 55.9&02 26 27&...&SA(s)a&11.89&1040&11.7&.51\\
NGC 4808&12 53 17.0&04 34 28&Sc(s)III&SAbc:&12.56:&766&10.2&-.68\\
\enddata
\label{tab1}
\tablecomments{
(1) Name of
galaxy, 
(2) and (3) Galaxy Coordinates (Epoch 1950.0), 
(4) Hubble types from BST or Sandage \& Tammann (1987) or 
Sandage \& Bedke (1994), 
(5) Hubble type from deVaucouleurs et al. (1991), 
(6) the total, face-on blue magnitude (B$_T^0$) from deVaucouleurs et al. (1991),
(7) the heliocentric radial velocity, 
(8) the projected angular distance in 
degrees of the galaxy from M87, and 
(9) the HI deficiency parameter, which was
calculated following the prescription of Giovanelli \& Haynes (1983).}
\end{deluxetable}

\begin{deluxetable}{lcc}
\tablecaption{Completeness By Hubble Type and B Magnitude in BST Region}
\tablewidth{0pt}
\tablehead{
\colhead{Type}& \colhead{B$_T^0$ $\le$ 12} & \colhead{12 $\le$ B$_T^0$ $\le$ 13}}
\startdata
S0-S0/a&29\% (4/14)&0\% (0/16)\\
Sa-Sab&83\% (5/6)&33\% (4/12)\\
Sb-Sbc&100\% (5/5)&100\% (2/2)\\
Sc-Scd&100\% (11/11) &55\% (11/20)\\
Sd-Sm& - (0/0) & 100\% (1/1)\\
Amorph& - (0/0) &100\%(2/2)\\
\enddata
\tablecomments{Completeness by Hubble type and total B magnitude for sample
galaxies within the BST survey region.}
\label{tab2}
\end{deluxetable}

\begin{deluxetable}{lllcclrrr}
\tablecaption{Virgo Cluster Observing Log}
\tablewidth{0pt}
\tablehead{
\colhead{(1)}& \colhead{(2)}&\colhead{(3)} &\colhead{(4)} &\colhead{(5)} &\colhead{(6)} & \colhead{(7)} & \colhead{(8)} & \colhead{(9)}\\
\colhead{Name}&
\colhead{Date}&
\colhead{Tel/Chip}&
\colhead{R Filt}&
\colhead{R Exp}&
\colhead{R X}&
\colhead{{FWHM}}&
\colhead{R $\sigma_b$}&
\colhead{R $\delta$s}\\
\colhead{}&
\colhead{}&
\colhead{}&
\colhead{H$\alpha$ Filt}&
\colhead{H$\alpha$ Exp}&
\colhead{H$\alpha$ X}&
\colhead{($^\prime$)}&
\colhead{H$\alpha$ $\sigma_b$}&
\colhead{H$\alpha$ $\delta$s}}
\startdata
NGC 4064&04/01/89 &KP9/TEK1 &nmR&420&1.38&1.6 &614 &110\\*
        &       &    &H$\alpha$1 &4000& 1.27&&44  &7\\
NGC 4178&03/23/87&KP9/TI2 &sR,R &3300 &1.60 & 3.0&197 &124\\*
& && H$\alpha$1&3900 &1.28 &&23 &7\\
NGC 4189&03/26/88 &KP9/TI2 &nmR &450 &1.49 &1.8 &362 &180\\*
& && H$\alpha$2&4500 &1.30 &&39 &14\\
NGC 4192&03/31/89 &KP9/TEK1 &nmR &420 &1.64 &2.5 &420 &504\\*
& & &H$\alpha$1 &4000 & 1.44& &45 &5\\
NGC 4206&04/02/89 &KP9/TEK1 &nmR &420 &1.88 &... &702 &...\\*
& & &H$\alpha$1 & 4000&1.59 & &67&...\\
NGC 4212&02/15/97 &WIYN/S2KB &R &180 &1.32 &2.6 &175 &75\\*
&03/22/87 &KP9/TI2 &sR &3000 &1.05 &3.1 &... &...\\*
& & &H$\alpha$1 & 3000&1.09 & &16 &5 \\
NGC 4216& 04/01/89&KP9/TEK1 &nmR &250 &1.14 &... &... &... \\*
& & & H$\alpha$1&2500 & 1.10&... &... &... \\
NGC 4237& 03/30/89&KP9/TEK1&nmR &420 &1.77 &2.2 &465 & 153\\*
& & & H$\alpha$1&4000 & 1.52&&47 & 8\\
NGC 4254&03/24/87 &KP9/TI2 &nmR &250 &1.13 &2.1 &300 &141 \\*
& & & H$\alpha$2&3000 &1.21 &&20 &8 \\
NGC 4293&03/25/88 &KP9/TI2 &nmR &300 &1.17  & 2.2&300 &219\\*
& & & H$\alpha$1&3600 &1.32 &&34 &41 \\
NGC 4294&03/31/89 &KP9/TEK1 &nmR &420 &1.26 &1.9 &169 & 132\\*
& & & H$\alpha$1&4000 &1.17 & &24 &8 \\
NGC 4298/&03/31/89 &KP9/TEK1 &nmR &420 &1.07 &1.8 &460 &110\\*
NGC 4302& & &H$\alpha$1&4000&1.06&&24 &8 \\
NGC 4299&03/30/89 &KP9/TEK1&nmR &420 &1.30 & 1.8 &153 &219 \\*
& & & H$\alpha$1&4000 &1.17 & &24 &7 \\
NGC 4303&03/29/88&KP9/TI2 &nmR &200 &1.25 &2.6 &880 &530 \\*
& & & H$\alpha$2&2000 &1.16 & &31 & 14\\
NGC 4321&02/09/99&CT9/TEK2K&R&2x300&1.45&1.5&380&127\\*
        &03/24/87&KP9/TI2 &nmR &230 &2.01 &2.0 &600 &... \\*
& & &H$\alpha$2 &2000 &1.59 & &47 &6 \\
NGC 4351&03/31/89 &KP9/TEK1 &nmR &420 &1.80 &2.4 &395 & 110\\*
& & &H$\alpha$2 &4000 &1.13 & &37 &8 \\
NGC 4380&04/01/89 &KP9/TEK1 &nmR &420 &1.10 &1.6 &307 & 110\\*
& & &H$\alpha$1 &4000 &1.90 & &42 & 8\\
NGC 4383/&02/04/95 &KP9/t2ka &R &3x240 &1.20 &1.4 &205 &43 \\*
UGC 7504&&&H$\alpha$4 &3x1200 &1.16 & &22 &9 \\
NGC 4394&03/30/89 &KP9/TEK1 &nmR &420 &1.05 &1.8 & 153 &153 \\*
& & &H$\alpha$1 &4000 &1.05 & &25 &8 \\
NGC 4405& 03/02/95&KP9/t2kA &R &2x450&1.04  &1.3 &170 &43\\*
        &       &&H$\alpha$4 &3x1800&1.05& &22 &9\\
NGC 4411B&04/02/89 &KP9/TEK1 &nmR&420 &1.20 &2.2 &220 & 220\\*
& & &H$\alpha$2 &4000 & 1.28& &29 & 8\\
NGC 4413&03/26/88 &KP9/TI2 &nmR &900 & 1.07&1.5 &120 &88 \\*
& & &H$\alpha$1 &4500 &1.08 & &22 & 7\\
NGC 4419 &02/15/97 &WIYN/S2KB &R &180 &1.30 & 1.9&156 &156 \\*
&03/26/87&KP9/TI2 &sR &600 &1.83 & 1.7&... &... \\*
& & &H$\alpha$1 &1200 &1.61 & &18 &4 \\
NGC 4424&02/02/95&KP9/t2ka &R&6x600 &1.15 &1.4 &93 &105 \\*
&04/02/89 &KP9/TEK1 &nmR &420 &1.4 &2.0 &658 &313 \\*
& & &H$\alpha$1 &400 &1.28 & &51 &13 \\
NGC 4429&02/05/95 & KP9/t2ka&R &4x135 &1.09 &1.1 &311& 108 \\*
& & &H$\alpha$3 &3x1200 &1.07 & &26 &11 \\
NGC 4438&04/02/89 &KP9/TEK1&nmR &420 &1.06 &2.2 &592 &219 \\*
& & &H$\alpha$2 &3000 &1.50 & &64 & 5\\
NGC 4450&04/01/89 &KP9/TEK1 &nmR&300 &1.06 &1.7 &395 & 164\\*
& & &H$\alpha$2 &3000 &1.11 & &42 &8 \\
 IC 3392&02/03/95 &KP9/t2ka &R &3x300&1.18&1.2&150 &43\\*
        &&&H$\alpha$4 &3x1800 &1.11& &20 &9\\
NGC 4457&03/03/93 &CT9/TEK1K-1 &R &300 &1.28 &1.9 &714 &320 \\*
& & &H$\alpha$ &3x2400 &1.22 & &38& 13\\
NGC 4459 & 03/30/89 &KP9/TEK1 &nmR &250 &1.06 &2.1 &395 & 219\\*
& & &H$\alpha$2 &3000 &1.11 & &33 & 20\\
NGC 4498&04/01/89 &KP9/TEK1 &nmR &420 &1.25 &1.6 &130 & 88\\*
& & &H$\alpha$2 &4000 &1.43 & &25 & 8\\
NGC 4501&03/29/88 &KP9/TI2 &nmR &250 &1.89 &2.6 &1266 & 527\\*
& & & H$\alpha$2&2000 &1.70 &&136 & 27\\
NGC 4519&03/26/88 &KP9/TI2 & nmR&600 &1.13 & 1.6&123 &105 \\*
& & &H$\alpha$2 &3600 &1.15 & & 25& 14\\
NGC 4522& 04/02/89&KP9/TEK1 &nmR&420 &1.16 &2.0 &153 &110 \\*
& & &H$\alpha$2 &4000 &1.12 & &24 & 5\\
NGC 4526&03/25/88 &KP9/TI2 & nmR&120&1.16 &2.0 &350 & 175\\*
& & &H$\alpha$1 & 2000&1.70 & &35 &40 \\
NGC 4527&03/25/87 &KP9/TI2 &nmR &450 &1.23 &1.9 &230 & 176\\*
& & &H$\alpha$2 &2500 &1.46 & &25 &8 \\
NGC 4532& 03/26/88&KP9/TI2 & nmR&900&1.23 & 1.7&97 & 176\\*
& & &H$\alpha$2 &3600 &1.15 & &25 &10\\
NGC 4535&03/29/88 &KP9/TI2 &nmR &300 &1.67 &2.2 &1406 & 176\\*
& & &H$\alpha$2 &2500 &1.46 & &115 & 20\\
NGC 4536&03/30/89 &KP9/TEK1&nmR  &400 &1.29 &1.8 &208 &219 \\*
& & &H$\alpha$2 &4000 &1.45 & &35 & 13\\
NGC 4548&03/29/88&KP9/TI2  &nmR &300 &1.50 & 3.6&770 &176 \\*
& & &H$\alpha$1 &2700 &1.06 & &81 & 16\\
NGC 4561& 03/26/87&KP9/TI2&nmR &500 &1.03 &2.1 &105 &53 \\*
& & & H$\alpha$2&6600 &1.31 & &16 & 7\\
NGC 4567&03/31/89&KP9/TEK1 &nmR &420 &1.23 &1.7 &147 & 219\\*
NGC 4568& & &H$\alpha$2 &4000 &1.40 & &24 & 8\\
NGC 4569&03/25/88 &KP9/TI2 &nmR&150 &1.15 &2.0  &240 &316 \\*
& & &H$\alpha$1 &3000 &1.10 & &41 & 68\\
NGC 4571&02/15/97&WIYN/s2kb &R &180 &1.09 &3.0 &156 & 150\\*
&03/23/87 &KP9/TI2 &sR &2500 &1.09 &2.2 &... &... \\*
& & &H$\alpha$1 &4500 &1.05 & &16 &7 \\
NGC 4579& 03/29/88&KP9/TI2 &nmR &250 &1.07 & 2.6&880 & 353\\*
& & &H$\alpha$2 &1500 &1.06 & &110 & 16\\
NGC 4580&02/02/95 &KP9/t2ka&R&3x450&1.23&1.4 &162 &43 \\*
& & &H$\alpha$3 &3x1800 &1.12 & &20 &11 \\
NGC 4586&04/09/92 &CT9/TEK1K &R &6x300 &1.31 &6.3 &300 &160\\*
& & &H$\alpha$5&4x1200 &1.28 & &... &... \\
NGC 4606/&02/06/95 &CT9/t2ka &R &3x240 &1.23 &1.4 &205 &65 \\*
NGC 4607 &&&H$\alpha$4 &4x900 &1.18 & &26 &11 \\
NGC 4639&03/31/89 &KP9/TEK1 &nmR &420 &1.66 &2.1 &79 &329 \\*
& & &H$\alpha$1 &4000 &2.26 & &14 &16 \\
NGC 4643&02/04/96 &CT9/TEK2K &R&2x300 &1.20 &1.5 &860 &450 \\*
& & &H$\alpha$6 &4x1350 &1.22& & 20 & 19\\
NGC 4647/&03/30/89 &KP9/TEK1 &nmR &350 &1.51 &1.9 &175 &329 \\*
NGC 4649&&&H$\alpha$2 &4000 &1.87 & &19 & 8\\
NGC 4651&02/15/97 &WIYN/s2kb & R&180 &1.19 &2.2 &156 &94 \\*
&03/23/87 &KP9/TI2 &sR &900 &2.01 &2.0 &... &... \\*
& & &H$\alpha$1 &2500 &1.68 & &16 &8 \\
NGC 4654&03/26/87 &KP9/TI2 &nmR&500 &2.14 &2.1 &141 & 176\\*
& & &H$\alpha$2 &4000 &1.74 & &24 & 11\\
NGC 4689&02/15/97 &WIYN/s2kb &R &180 &1.23 &2.0 &230 &125 \\*
&03/25/87 &KP/TI2 &sR &2500&1.06 &2.2 &... & ...\\*
& & &H$\alpha$2 &4000 &1.29 & &18 &7 \\
NGC 4694&02/04/95 &KP9/t2ka &R &2x450 &1.12 &1.5 &173 &43 \\*
& & &H$\alpha$3 &3x1200 &1.09 & &17 &11 \\
NGC 4698&02/05/95 &KP9/t2ka &R& 3x180 &1.34 &1.2 &280 &65 \\*
& & &H$\alpha$3 &3x900 &1.26 & &15 &9 \\
NGC 4710&03/26/87 &KP9/TI2 &nmR&300 &1.36 &1.7 &510&175 \\*
& & &H$\alpha$2 &1850 &1.11 & &40 &41 \\
NGC 4713&03/26/87 &KP9/TI2 &nmR&800 &1.18 &2.0 &79 &108 \\*
& & &H$\alpha$1 &3800 &1.12 & &19 &11 \\
NGC 4772&02/25/93 &CT9/TEK2K &R &1x900 &1.33 &2.4 &446 &128 \\*
& & & H$\alpha$5&2x2400 &1.19 & &51 &16 \\
NGC 4808&02/15/97  &WIYN/s2kb&R &180 &1.20 &2.5 &160 &94 \\*
&03/23/87 &KP9/TI2 &sR &1500 &1.20 &2.1 &... &... \\*
& & &H$\alpha$1 &4000 &1.27 & &18 &7 \\
\enddata
\label{obslog}
\tablecomments{
(1) name of the galaxy, 
(2) the date of the observation, 
(3) the telescope
and chip (where chip characteristics are listed in Table~\ref{chip}, and
KP9 means Kitt Peak 0.9 m and CT9 means CTIO 0.9m), 
(4) filter code
(listed in Table~\ref{filter}), 
(5) exposure time in seconds,
(6) airmass of the observation, 
(7) the full width half maximum (FWHM) in arcseconds of the processed images 
(the R and H$\alpha$ images were convolved to the same seeing in the 
reduction process), 
(8) the sky background sigma, 
(9) and the estimated uncertainty in the sky background
level. The units of columns (8) and (9) are 
$10^{-18}$ erg cm$^{-2}$ s$^{-1}$ arcsec$^{-2}$.
R flux values may be converted to magnitudes using a zeropoint of 13.945.}
\end{deluxetable}

\twocolumn

\begin{deluxetable}{lcrr}
\tablecaption{Chip Codes and Characteristics}
\tablewidth{0 pt}
\tablehead{
\colhead{Detector} & \colhead{Scale} & \colhead{Size} & \colhead{FOV}\\
\colhead{} & \colhead{$^{(\prime\prime}$ pix$^{-1}$)} & \colhead{(pix)} & \colhead{$^{(\prime}$)}
}
\startdata
TI2&0.86&396&5.7\\
TEK1&0.77&512&6.6\\
t2ka&0.68&2048&23.2\\
TEK2K&0.40&2048&13.7\\
TEK1K&0.40&1024&6.8\\
s2kb&0.20&2048&6.8\\ 
\enddata
\label{chip}
\end{deluxetable}

\begin{deluxetable}{llr}
\tablecaption{Filter Characteristics}
\tablewidth{0 pt}
\tablehead{
\colhead{Filter} & \colhead{$\lambda_{cent}$} & \colhead{$\delta\lambda$}\\
\colhead{} & \colhead{(\AA)} & \colhead{(\AA)}} 
\startdata
H$\alpha$1&6563&80\\
H$\alpha$2&6608&76\\
H$\alpha$3&6573&68\\
H$\alpha$4&6618&74\\
H$\alpha$5&6563&78 \\
H$\alpha$6&6606&75\\
R&6425&1540\\
nmR&6470&1110 \\
sR&7024 &380\\
\enddata
\label{filter}
\end{deluxetable}

\onecolumn

\begin{deluxetable}{lrrrrrrrrr}
\tablecaption{Parameters and Derived Quantities for the R Surface Photometry}
\tablewidth{0pt}
\tablehead{
\colhead{(1)}& \colhead{(2)}&\colhead{(3)} &\colhead{(4)} &\colhead{(5)} &\colhead{(6)} & \colhead{(7)} & \colhead{(8)} & \colhead{(9)} &\colhead{10)} \\
\colhead{Name}&
\colhead{Inc}&
\colhead{PA}&
\colhead{r$_{24}$}&
\colhead{R$_{24}$}&
\colhead{r$_{25}$}&
\colhead{R$_{25}$}&
\colhead{C30}&
\colhead{r$_d$}&
\colhead{$\Delta$ r$_d$}\\
\colhead{}&
\colhead{($^{\circ}$)}&
\colhead{($^{\circ}$)}&
\colhead{($^{\prime\prime}$)}&
\colhead{(mag)}&
\colhead{($^{\prime\prime}$)}&
\colhead{(mag)}&
\colhead{}&
\colhead{($^{\prime\prime}$)}&
\colhead{($^{\prime\prime}$)}
}
\startdata
NGC 4064&0.391 (70) &150&104&11.2&138&11.1&0.43&44&2\\ 
NGC 4178&0.342 (74) &33&119&11.4&146&11.2&0.25&45&4\\ 
NGC 4189&0.695 (47)\tablenotemark{a}&70 \tablenotemark{b}&72&11.5&86&11.4&0.29&20&5\\ 
NGC 4192&0.276 (79)\tablenotemark{c} &155&218&9.7&260&9.6&0.34&72&4\\ 
NGC 4212&0.643 (51)\tablenotemark{d}&75&89&10.7&104&10.6&0.39&22&4\\ 
NGC 4237&0.682 (48) &108&69&11.2&92&11.2&0.47&16&4\\ 
NGC 4254&0.839 (34) &57:&150&9.4&178&9.4&0.46&33&3\\  
NGC 4293&0.438 (67) &66\tablenotemark{a}&162&10.1&216&10.0&0.40&60&5\\ 
NGC 4294&0.391 (70) &155&74&11.8&92&11.7&0.33&22&2\\ 
NGC 4298&0.530: (60)&135&109&10.8&153&10.8&0.36&32&3\\ 
NGC 4299&0.927: (22)\tablenotemark{e} &96\tablenotemark{f}&52&12.2&0&0.0&0.33&15&4\\ 
NGC 4303&0.906 (26) &42&177&9.2&0&0.0&0.51&42&3\\ 
NGC 4321&0.875 (30)\tablenotemark{f}&153\tablenotemark{f}&239&9.1&319&9.0&0.40&75&4\\
NGC 4351&0.695 (47)\tablenotemark{a}&80&62&12.3&92&12.1&0.37&25&5\\ 
NGC 4380&0.530 (60) &158&104&11.1&129&11.0&0.33&35&5\\ 
NGC 4383&0.643 (51)\tablenotemark{a}&15\tablenotemark{a}&61&11.7&87&11.6&0.63&...&...\\ 
NGC 4394&0.906\tablenotemark{e} (26) &108\tablenotemark{e}&112&10.5&145&10.5&0.47&33&2\\ 
NGC 4405&0.695 (47)\tablenotemark{a}&15&54&11.8&69&11.8&0.47&18&5\\ 
NGC 4411B&1.00 (0)\tablenotemark{e} &0:\tablenotemark{e}&80&12.1&99&12.0&0.32&29&2\\ 
NGC 4413&0.656 (50) & 60& 65& 11.9& 86& 11.9&  0.37& 20&2\\ 
NGC 4419& 0.391 (70)& 133& 84& 10.7& 103& 10.6&  0.52& 20&2\\ 
NGC 4424&0.616 (54)\tablenotemark{a}& 90& 93& 11.2& 125& 11.1&  0.42& 33&4\\
NGC 4429&0.500 (62) &90:\tablenotemark{a}& 171& 9.7 &220& 9.7& 0.54& 57&4\\ 
NGC 4450&0.707 (46) & 175& 166& 9.6& 205& 9.5& 0.49& 41&4\\ 
 IC 3392&0.454 (65)& 37& 66 &11.8& 91& 11.8& 0.37& 22&5\\ 
NGC 4457&1.00: (0)\tablenotemark{e} & 0:\tablenotemark{e}& 128& 10.1& 0& 0.0& 0.66& 40&3\\ 
NGC 4459&0.819 (36)&110& 138 &10.0& 177& 10.0&  0.71& 40&3\\ 
NGC 4498&0.485 (63)& 136& 75& 12.0& 95& 11.9& 0.31& 25&2\\ 
NGC 4501&0.50 (62)& 140& 185& 9.1& 0& 0.0& 0.45& 46&3\\ 
NGC 4519&0.743 (43)& 146& 73& 11.8& 85& 11.8& 0.35& 21&4\\ 
NGC 4522&0.276 (79) & 36& 78& 12.2& 107& 12.0&  0.28& 30&3\\ 
NGC 4527&0.309 (76)& 67& 157& 10.1& 214& 10.0& 0.42& 50&5\\ 
NGC 4532&0.391 (70)\tablenotemark{g}& 160& 66& 11.7& 89& 11.7& 0.39& 18&3\\ 
NGC 4535&0.731 (44)& 0& 174& 9.8& 0& 0.0& 0.29& 60&6\\ 
NGC 4536&0.391 (70)& 125& 190& 10.1& 231& 10.0& 0.36& 70&7\\ 
NGC 4548&0.809 (37)& 136\tablenotemark{f}& 162& 9.6& 191& 9.6& 0.43& 50&5\\ 
NGC 4561&0.829 (35)& 35& 46& 12.5& 63& 12.4&  0.39& 18&4\\ 
NGC 4567&0.695 (47)& 85& 113& 10.9& 141& 10.9& 0.50& 31&3\\ 
NGC 4568&0.438 (67)& 23& 128& 10.5& 165& 10.4& 0.44& 36&3\\ 
NGC 4569&0.469 (64)&23&209&9.3&266&9.3&0.43&66&4\\ 
NGC 4571&0.829 (35)\tablenotemark{a}&40\tablenotemark{a}&120&10.8&152&10.7&0.34&40&4\\
NGC 4579&0.799 (38)&95&166&9.2&0&0.0&0.52&42&3\\ 
NGC 4580&0.719 (45)&158\tablenotemark{a}&72&11.3&87&11.3&0.41&19&5\\ 
NGC 4586&0.342 (74)&114&129&11.4&165&11.3&0.56&...&...\\ 
NGC 4606&0.438 (67)&40\tablenotemark{a}&76&11.8&103&11.7&0.42&25&5\\ 
NGC 4639&0.669 (49)&123&79&11.2&100&11.1&0.50&20&2\\ 
NGC 4643&0.777 (40)&30\tablenotemark{a}&141&10.0&0&0.0&0.60&45&4\\ 
NGC 4647&0.809 (37)&100\tablenotemark{a}&86&10.9&0&0.0&0.40&25&3\\ 
NGC 4651&0.642 (51)&71\tablenotemark{a} \tablenotemark{f}&115&10.3&157&10.2&0.55&26&5\\ 
NGC 4654&0.629 (52)&125&119&10.5&158&10.5&0.35&36&3\\ 
NGC 4689&0.819 (36)&166&143&10.4&195&10.3&0.38&50&5\\ 
NGC 4694&0.755 (42)&140&85&11.3&120&11.2&0.62&40&3\\
NGC 4698&0.500 (62)\tablenotemark{a}&168&127&10.3&187&10.2&0.51&70&5\\ 
NGC 4713&0.656 (50)&95&68&11.6&78&11.6&0.35&17&4\\ 
NGC 4772&0.500 (62)&145&117&10.9&144&10.9&0.51&42&4\\
NGC 4808&0.391 (70)&127&69&11.5&83&11.4&0.38&19&4\\
\enddata
\tablecomments{
(1) Name of galaxy,
(2) axial ratio and, in parentheses, inclination calculated using the 
Hubble (1926) conversion (see Section~\ref{sppar}, 
(3) position angle, 
(4) r$_{24}$, the radius in units of arcsecs at the 24 R mag/arcsec$^2$ 
isophote, (5) the magnitude within r$_{24}$, (6)r$_{25}$, the radius in 
units of arcsecs at the 25 R mag/arcsec$^2$ isophote, (7) the magnitude 
within r$_{25}$, (8) the central R light
concentration parameter, (9) the disk scalelength, and (10) the uncertainty
in the disk scalelength.}
\label{rtab}

\tablenotetext{a}{Pierce \& Tully (1988) inclination of 
55$^{\circ}$ not supported by outer isophotes} 
\tablenotetext{b}{Value different by more than 5$^\circ$ from RC3.}
\tablenotetext{c}{Pierce \& Tully (1988) inclination of 
86$^{\circ}$ not supported by outer isophotes; Warmels (1988) HI analysis indicates an inclination of 70$^{\circ}$.}
\tablenotetext{d}{Pierce \& Tully (1988) inclination of
44$^{\circ}$ not supported by outer isophotes}
\tablenotetext{e}{Galaxy is close to face-on, so the inclination and
position angle are less certain.}
\tablenotetext{f}{Value taken from Warmels (1988).}
\tablenotetext{g}{Pierce \& Tully (1988) inclination of 
61$^{\circ}$ not supported by outer isophotes} 
\end{deluxetable}

\begin{deluxetable}{lcrrrccc}
\tablecaption{Parameters and Derived Quantities  from the H$\alpha$ surface photometry}
\tablewidth{0pt}
\tablehead{
\colhead{(1)}& \colhead{(2)}&\colhead{(3)} &\colhead{(4)} &\colhead{(5)} &\colhead{(6)} & \colhead{(7)} & \colhead{(8)} \\
\colhead{Name}&
\colhead{log F$_{H\alpha}$}&
\colhead{r$_{HII}$}&
\colhead{r$_{H\alpha95}$}&
\colhead{r$_{H\alpha17}$}&
\colhead{log F$_{H\alpha17}$}&
\colhead{CH$\alpha$}&
\colhead{$\rm \frac{CH\alpha}{C30}$}\\
\colhead{}&
\colhead{}&
\colhead{($^{\prime\prime}$)}&
\colhead{($^{\prime\prime}$)}&
\colhead{($^{\prime\prime}$)}&
\colhead{}&
\colhead{}&
\colhead{}}
\startdata
NGC 4064& -12.31&40&25&25&-12.33&0.97& 2.26\\
NGC 4178&-11.67&143&120&119&-11.69&0.15&0.60\\
NGC 4189&-11.92&71&58&63&-11.93&0.15&0.52\\
NGC 4192&-11.50&243&194&144&-11.61&0.17&0.50\\
NGC 4212&-11.77&71&55&63&-11.78&0.37&0.95\\
NGC 4237&-12.25&43&44&40&-12.25&0.55&1.17\\
NGC 4254&-10.95&168&123&142&-10.95&0.32&0.70\\
NGC 4293&-12.64&48&44&48&-12.64&1.00&2.50\\
NGC 4294&-11.79&69&51&60&-11.79&0.42&1.27\\
NGC 4298&-11.95&86&71&76&-11.96&0.45&1.25\\
NGC 4299&-11.70&40&28&39&-11.70&0.46&1.39\\
NGC 4303&-10.76&191&133&166&-10.76&0.49&0.96\\
NGC 4321&-11.13&228&148&159&-11.14&0.46&1.15\\
NGC 4351&-12.47&39&30&33&-12.48&0.77&2.08\\
NGC 4380&-12.64&72&63&33&-12.93&0.49&1.48\\
NGC 4383&-11.56&154&91&53&-11.62&0.84&1.33\\
NGC 4394&-12.25&100&89&59&-12.44&0.25&0.53\\
NGC 4405&-12.47&29&19&22&-12.48&0.91&1.94\\
NGC 4411B&-12.28&79&76&61&-12.33&0.35&1.09\\
NGC 4413&-12.31&39&34&39&-12.31&0.58&1.57\\
NGC 4419&-12.70&52&46&32&-12.83&0.64&1.23\\
NGC 4424&-12.22&40&27&32&-12.23&1.00&2.38\\
NGC 4429&-13.15&...&...&...&...&...&...\\
NGC 4438&-11.77&...&...&...&...&...&...\\
NGC 4450&-12.30&111&92&61&-12.38&0.69&1.41\\
 IC 3392&-12.60&40&30&25&-12.63&0.82&2.22\\
NGC 4457&-11.91&38&33&...&-11.91&1.00&1.52\\
NGC 4459&-12.32&...&...&...&...&...&...\\
NGC 4498&-12.06&90&70&71&-12.08&0.25&0.81\\
NGC 4501&-11.21&140&112&132&-11.21&0.44&1.02\\
NGC 4519&-11.75&109&86&75&-11.79&0.14&0.41\\
NGC 4522&-12.39&149&54&43&-12.43&0.56&2.00\\
NGC 4527&-11.35&172&138&128&-11.38&0.29&0.69\\
NGC 4532&-11.40&119&68&72&-11.41&0.82&2.10\\
NGC 4535&-11.21&186&156&168&-11.22&0.17&0.59\\
NGC 4536&-11.42&256&198&134&-11.51&0.46&1.28\\
NGC 4548&-11.96&141&126&107&-12.06&0.11&0.26\\
NGC 4561&-12.11&42&34&40&-12.11&0.30&0.77\\
NGC 4567&-11.91&59&41&54&-11.91&0.55&1.10\\
NGC 4568&-11.59&99&72&79&-11.60&0.62&1.41\\
NGC 4569&-11.83&72&65&72&-11.83&0.92&2.14\\
NGC 4571&-11.97&109&92&86&-12.01&0.27&0.79\\
NGC 4579&-11.54&108&90&99&-11.55&0.61&1.17\\
NGC 4580&-12.57&28&23&26&-12.58&0.93&2.27\\
NGC 4606&-12.67&50&47&28&-12.86&0.57&1.90\\
NGC 4639&-11.94&82&67&70&-11.95&0.25&0.50\\
NGC 4643&-12.30&...&...&...&...&...&...\\
NGC 4647&-11.72&80&57&62&-11.73&0.44&1.10\\
NGC 4651&-11.51&144&92&101&-11.53&0.56&1.02\\
NGC 4654&-11.60&176&101&92&-11.62&0.30&0.83\\
NGC 4689&-11.92&81&64&70&-11.93&0.67&1.76\\
NGC 4694&-12.87&30&25&20&-12.92&0.95&1.53\\
NGC 4698&-12.28&267&247&22&-13.24&0.31&0.61\\
NGC 4713&-11.62&101&63&72&-11.63&0.28&0.80\\
NGC 4772&-12.36&79&72&57&-12.47&0.43&0.84\\
NGC 4808&-11.71&129&65&67&-11.73&0.24&0.63\\
\enddata
\tablecomments{
(1) Name of galaxy, 
(2) log of the total H$\alpha$ flux in units of 
erg cm$^{-2}$ s$^{-1}$, 
(3) radius in arcseconds of the outermost HII region,
(4) radius in arcseconds which contains 95\% of the total H$\alpha$ flux, 
(5) radius in arcseconds at which the surface brightness falls to 
17 x 10$^{-18}$ erg cm$^{-2}$ s$^{-1}$ arcsec$^{-1}$,
(6) log of the H$\alpha$ flux within r$_{H\alpha17}$ 
in units of erg cm$^{-2}$ s$^{-1}$, (7) the
H$\alpha$ concentration, and (8) the H$\alpha$ concentration
normalized by the central R concentration.}
\label{hatab}
\end{deluxetable}

\end{document}